\documentclass[english,aps,prd,superscriptaddress,preprintnumbers,nofootinbib,preprint]{revtex4-1}
\usepackage[T1]{fontenc}
\usepackage[utf8]{inputenc}
\usepackage{verbatim}
\usepackage{empheq}
\usepackage{amssymb}
\usepackage{amsmath}
\usepackage{amsfonts}

\usepackage[colorlinks]{hyperref}
\hypersetup{
 citecolor=blue,
 linkcolor=blue,
 urlcolor=blue}

\usepackage{xcolor}

\usepackage{dcolumn}
\usepackage{graphicx}
\usepackage{etoolbox} 

\newcommand{\dq}[1]{[\textrm{d}p_{#1}]}

\newcommand{\ep}{\epsilon}
\newcommand{\as}{\alpha_s}
\newcommand{\sh}{s_{\rm had}}

\newcommand{\bb}{ {b\bar{b}} }

\newcommand{\mev}{\rm MeV}
\newcommand{\gev}{\rm GeV}

\newcommand{\eqspace}{\hphantom{{}={}}}
\newcommand{\msq}[2]{\left|\mathcal{M}{\left(#1 \,;\, #2\right)}\right|^2}


\usepackage{babel}
\usepackage{color}
\begin{document}
\global\long\def\order#1{\mathcal{O}\left(#1\right)}
\global\long\def\d{\mathrm{d}}
\global\long\def\P{P}
\global\long\def\amp{{\mathcal M}}
\preprint{OUTP-21-23P}
\preprint{TTP21-032}
\preprint{P3H-21-066}
\preprint{CERN-TH-2021-146}
\preprint{TIF-UNIMI-2021-15}

\def\BNL{ High Energy Theory Group, Physics Department, Brookhaven
  National Laboratory, Upton, NY 11973, USA }

\def\KIT{ Institute for Theoretical Particle Physics, KIT, Karlsruhe,
  Germany }
\def\IKP{ Institute for Astroparticle Physics, KIT,
  Karlsruhe, Germany }
\def\CERN{ Theoretical Physics Department,
  CERN, 1211 Geneva 23, Switzerland }
\def\OX{ Rudolf Peierls Centre
  for Theoretical Physics, Clarendon Laboratory, Parks Road, Oxford
  OX1 3PU, UK and Wadham College, Oxford OX1 3PN, UK }
\def\MIL{ Tif Lab, Dipartimento di Fisica, Università di Milano and INFN, Sezione di Milano, Via 
  Celoria 16, I-20133 Milano, Italy}

\title{ NNLO QCD corrections to weak boson fusion Higgs boson production
  in the $H \to b\bar b$ and $H \to WW^* \to 4l$ decay channels} 

\author{Konstantin~Asteriadis}
\email[Electronic address: ]{kasteriad@bnl.gov}
\affiliation{\BNL}

\author{Fabrizio~Caola}
\email[Electronic address: ]{fabrizio.caola@physics.ox.ac.uk}
\affiliation{\OX}

\author{Kirill~Melnikov}
\email[Electronic address: ]{kirill.melnikov@kit.edu}
\affiliation{\KIT}

\author{Raoul~R\"ontsch }
\email[Electronic address: ]{raoul.rontsch@cern.ch}
\affiliation{\CERN}
\affiliation{\MIL}

\begin{abstract}
  We compute the next-to-next-to-leading order QCD corrections to
  Higgs boson production in weak boson fusion followed by its decay to
  a $\bb$ pair or to a pair of leptonically-decaying $W$ bosons.
  Our calculation
allows us to compute realistic fiducial cross sections and assess the
impact of fiducial cuts applied to the Higgs boson decay products on
the magnitude of QCD radiative corrections in weak boson fusion.
\end{abstract}

\maketitle

\section{Introduction}
 \label{sec:intro}
 Precision studies of Higgs boson properties are central to the
 physics program of the Run III
 and high-luminosity phases of the LHC.  Currently, all major Higgs
 production cross sections and decay rates are known experimentally to
 a precision of about twenty percent or
 better~\cite{ATLAS:2019nkf,CMS:2018uag}.  These measurements are 
 used to determine Higgs couplings to a variety of elementary
 particles, confirming that the Higgs boson profile emerging from the
 LHC data is very consistent with expectations based on the Standard
 Model.

Further exploration of the Higgs boson will lead to an even better
understanding of its properties. Central to this endeavor is the
overarching goal of the LHC experiments to determine Higgs couplings
with a few percent precision, allowing for a detailed exploration  of the
structure of the Standard Model.  To facilitate this progress,
precise theoretical predictions for all the major Higgs
boson production and decay processes are required. Such predictions
must, on the one hand, account for higher order radiative corrections
and, on the other hand, describe observable final states in as much
detail as possible. The recent past has seen impressive progress in
the development of high-quality theoretical predictions for Higgs
production and decay processes at colliders, see e.g.~Ref.~\cite{Heinrich:2020ybq} for a review.

In this paper, we focus on Higgs boson production in weak boson fusion (WBF).
Being  the channel with the next-to-largest cross section at
the LHC, it allows for detailed studies of the structure of the Higgs 
sector. Indeed, phenomenologically,  this channel is important
for a direct determination  of the Higgs couplings to the
electroweak bosons, for investigating   the CP-structure of the Higgs
boson~\cite{Plehn:2001nj,ATLAS:2016ifi,CMS:2019jdw} and for studies
of Higgs decays into invisible particles~\cite{ATLAS:2015gvj,CMS:2018yfx}.

Higgs production in WBF has been investigated by both ATLAS and CMS using a
number of Higgs decay modes. One finds
$\mu_{\rm VBF} = 1.21 \pm
0.18~{\rm(stat.)}\pm 0.15~{\rm(syst.)}$ (ATLAS)~\cite{ATLAS:2019nkf} and $\mu_{\rm VBF} = 0.73 \pm
0.23~{\rm(stat.)}\pm 0.16~{\rm(syst.)}$ (CMS)~\cite{CMS:2018uag}  for the  signal strength in this channel
relative to the Standard Model expectations.
These studies are supported by the development of  sophisticated theoretical 
tools that allow accurate descriptions of  Higgs boson production in WBF.\footnote{For a review of the
  state-of-the-art predictions for this channel, see Ref.~\cite{LHCHiggsCrossSectionWorkingGroup:2016ypw}.
  Recent phenomenological studies are summarized in  Ref.~\cite{Buckley:2021gfw}.}

At leading order, WBF production involves the two incoming partons emitting space-like
$Z$'s and $W$'s that fuse
into the Higgs boson.  This gives WBF events a characteristic
signature of two forward jets in opposite hemispheres, and means that
WBF can be viewed as a double deep inelastic scattering (DIS)
process~\cite{Han:1992hr}. At next-to-leading order (NLO) in QCD,
color conservation forbids interactions that connect the two quark
lines, so that the double-DIS picture of WBF is still
exact.\footnote{This statement does not hold for interference
  contributions to the $qq \to Hqq$ amplitude squared. However, these contributions are known to be   tiny when WBF cuts are
  applied~\cite{Ciccolini:2007jr,Ciccolini:2007ec,Buckley:2021gfw}.} As
a consequence of this simplification, NLO QCD corrections to WBF
were computed early on and have been known for almost twenty
years by now~\cite{Figy:2003nv}.

At
next-to-next-to-leading order (NNLO), QCD interactions between two 
quark lines become possible, so that the double-DIS picture of
WBF is no longer exact. However, the double-DIS and the so-called
non-factorizable contributions are separately finite and
gauge-invariant and, therefore,  can be studied independently.
Theoretical predictions for the double-DIS  contributions are very advanced. In this approximation, the   
total cross section is known to NNLO~\cite{Bolzoni:2010xr,*Bolzoni:2011cu}
and N$^3$LO~\cite{Dreyer:2016oyx} in QCD, while fully differential
results -- which are crucial for a reliable modeling of the WBF  process
-- are available at NNLO QCD~\cite{Cacciari:2015jma,Cruz-Martinez:2018rod}.

On the contrary,  the non-factorizable contributions are much less understood since no
exact results exist in this case. Recently, the leading
non-factorizable QCD corrections that appear at NNLO QCD have been
estimated~\cite{Liu:2019tuy}.  Although non-factorizable corrections
are known to be color-suppressed \cite{Bolzoni:2010xr,*Bolzoni:2011cu},
it was explicitly shown in Ref.~\cite{Liu:2019tuy} that they get
enhanced by additional factors of $\pi^2$ leading to a partial
compensation of the color suppression factor.
A comprehensive discussion of  phenomenological aspects of  non-factorizable corrections can be
found in Ref.~\cite{Dreyer:2020urf}.

All the NNLO calculations mentioned above do not include the decay of
the Higgs boson  and, therefore, cannot describe realistic final
states in WBF. Since the Higgs boson is a narrow scalar particle, its production
and decay stages are completely separated. Hence, the 
inclusion of Higgs decays is, in principle, straightforward. However, in practice this turns out to be 
non-trivial for fully differential NNLO QCD calculations. One may
naively think of performing a NNLO calculation with a stable Higgs boson, storing 
events with  the relevant kinematic information and including the Higgs decays in a
second stage. In fact, this strategy is
routinely employed in  complex  NLO calculations.  However, the number of
events needed for a NNLO calculation of the WBF type is extremely high, making this
approach impractical.\footnote{A recent summary on progress in
  this direction was given in  Ref.~\cite{Maitre:2020blv}.} As the  result, NNLO QCD
predictions for Higgs production in WBF including Higgs decays are 
currently not available.
This limitation is important as kinematic cuts applied to the
decay products of the Higgs boson may alter the impact of NNLO QCD
corrections.  Even if such  modifications turn out to be  small, they may still be
relevant at NNLO QCD accuracy, given that both the size of the NNLO
corrections and the residual scale uncertainty for Higgs production in WBF are  at the level
of a few percent. 

In this paper, we make a first step towards  NNLO
QCD predictions for Higgs production in WBF with realistic final
states. Similarly to the earlier calculations of
Ref.~\cite{Cacciari:2015jma,Cruz-Martinez:2018rod} we only consider
the double-DIS contributions but we include Higgs decays in our computation. 
We consider two representative cases -- Higgs decays to a
pair of $b$ quarks and Higgs decays to two $W$ bosons which decay
leptonically in turn.  Since both of these final states have large branching ratios,  they 
have been extensively studied by the
ATLAS~\cite{ATLAS:2020bhl,ATLAS:2014aga,*ATLAS:2016mzy,*ATLAS:2018jvf,*ATLAS:2018xbv,*ATLAS:2020cvh}
and CMS~\cite{CMS:2018zzl,CMS:2015ebl,*CMS:2020dvg} collaborations.   Our calculations allow us to
  investigate the impact of kinematic constraints applied to final state particles on
  the magnitude of QCD corrections to  fiducial cross sections and
  kinematic distributions in the presence of realistic selection criteria.

In principle, a proper modeling of Higgs-boson decays to $b$ pairs
would require the inclusion of $b$-quark mass effects (so that a realistic
jet algorithm can be adopted) and higher-order QCD radiative
corrections. Although all the required ingredients exist
\cite{Bernreuther:2018ynm,Behring:2020uzq,Behring:2020uzq}, combining
them into an efficiently working computer program is a major
undertaking if NNLO QCD precision is desired.  For this reason, in
this paper we restrict ourselves to Higgs decays into massless $b \bar
b$ pairs at leading order in QCD.\footnote{Of course, for leading order $H \to b \bar b$
  decay dealing with massless, as opposed to massive,  $b$-quarks does not change the 
  complexity of the calculation.}   This is the first non-trivial step
towards a NNLO-accurate description of the weak boson fusion process
with Higgs decays into $b \bar b$ pairs.

On a technical side, we note that the previous NNLO QCD calculations
of Refs.~\cite{Cacciari:2015jma,Cruz-Martinez:2018rod} were performed
using the projection-to-Born~\cite{Cacciari:2015jma} and
antenna subtraction~\cite{Gehrmann-DeRidder:2005btv,*Gehrmann-DeRidder:2005svg,
  *Gehrmann-DeRidder:2005alt,*Daleo:2006xa,*Daleo:2009yj,
  *Boughezal:2010mc,*Gehrmann:2011wi,*Gehrmann-DeRidder:2012too,
  *Currie:2013vh} methods,
respectively, to regulate infrared singularities.  For the computation
described in this paper we employ the  nested soft-collinear
subtraction scheme~\cite{Caola:2017dug}, using analytic formulas for
the integrated subtraction terms derived in
Ref.~\cite{Asteriadis:2019dte}.  This is the first phenomenological
application of this method to a complex LHC process involving
final-state jets, marking  an important step in its development.

The paper is organized as follows. In Section~\ref{sect2} we briefly summarize some 
technical aspects of the calculation of NNLO QCD corrections to Higgs production in WBF
within the nested soft-collinear subtraction scheme. In
Section~\ref{sect3} we apply our computation to perform NNLO QCD
phenomenological studies at the 13 TeV LHC. We first present results
for stable Higgs boson  (Section \ref{sect3a}), showing perfect agreement
with earlier calculations. We then discuss the $H\to b\bar b$
(Section \ref{sect3b}) and $H\to WW^*\to \ell^- \bar\nu \ell^+\nu$
(Section \ref{sect3c}) cases, focusing on the interplay of fiducial cuts on
the final state particles and NNLO QCD corrections.
We conclude in Section~\ref{sect4}.

\section{Nested soft-collinear subtraction calculation of NNLO QCD corrections
  to weak boson fusion}
\label{sect2}
 
In this section we summarize the technical aspects of the calculation
including a brief discussion of the nested soft-collinear subtraction
scheme \cite{Caola:2017dug} and an explanation of how we apply it to
the computation of factorizable NNLO QCD corrections to Higgs boson
production in weak boson fusion. We note that  this section is not meant to provide a self-contained
discussion of the nested soft-collinear subtraction scheme. We refer the interested reader
to Refs.~\cite{Caola:2017dug,Asteriadis:2019dte} for a thorough explanation
of all the relevant details.

It is well-known that fully-differential QCD computations suffer from
infra-red and collinear singularities that appear differently in
virtual and real corrections. In the case of virtual corrections, these
singularities manifest themselves as explicit poles in the
dimensional regularization parameter.  The universal structure of
these singularities has been known for a long time~\cite{Catani:1998bh,Becher:2009cu,*Becher:2009qa}.

The situation with the real-emission contributions is quite different.
Indeed, in this case infra-red and collinear singularities arise from
the integration over the phase space of final state partons that
appear at
higher orders in QCD, in addition to the particles present in the Born process. However, if one's goal is a fully-differential
computation valid for arbitrary infra-red safe observables, such
integration is not possible both in theory and in practice.

The solution to this problem involves isolating phase-space regions
that contribute to infra-red and collinear singularities. 
The standard ways for dealing with them are the so-called slicing and subtraction
methods, which regulate the singular regions of the phase space by
adding and subtracting cleverly-constructed counterterms.
In the past decade, a large number of such  schemes has
been developed for NNLO QCD
computations~\cite{Gehrmann-DeRidder:2005btv,*Gehrmann-DeRidder:2005svg,
  *Gehrmann-DeRidder:2005alt,*Daleo:2006xa,*Daleo:2009yj,*Boughezal:2010mc,
  *Gehrmann:2011wi,*Gehrmann-DeRidder:2012too,*Currie:2013vh,
  Czakon:2010td,*Czakon:2011ve,*Czakon:2014oma,*Boughezal:2011jf,
  Cacciari:2015jma,Catani:2007vq,*Grazzini:2008tf,
  Gaunt:2015pea,*Boughezal:2015dva,*Campbell:2017hsw,
  DelDuca:2016csb,*DelDuca:2016ily,
  Magnea:2018hab,*Magnea:2018ebr,
  Herzog:2018ily}.
In this paper, we employ the so-called nested soft-collinear
subtraction scheme \cite{Caola:2017dug}.

The idea behind this scheme is an iterative 
construction of subtraction counterterms, starting from the soft
ones. The soft subtraction terms are given by universal eikonal
currents multiplied by amplitudes with lower multiplicities. For the
non-trivial case of double-unresolved contributions, they were
integrated over the phase space of unresolved partons in
Ref.~\cite{Caola:2018pxp} using the universal double-soft limits of
scattering amplitudes computed in Ref.~\cite{Catani:1999ss}.

The collinear regularization procedure is then applied to
soft-regulated matrix elements. In analogy with the FKS construction
at NLO~\cite{Frixione:1995ms,*Frixione:1997np}, different collinear
directions are separated with the help of suitably-constructed
partitions of unity. For triple-collinear configurations, different
strongly-ordered collinear limits can be approached in different ways; as the result, 
 an angular ordering is employed to uniquely define them. This can
be achieved by introducing a particular phase-space parametrization
for the two unresolved partons \cite{Czakon:2010td,*Czakon:2011ve}
that allows for a natural separation of various collinear configurations.
Among the various collinear subtraction terms that need to be
considered, the triple-collinear ones are the most complicated; they
were integrated over the phase space of the unresolved partons in
Ref.~\cite{Delto:2019asp}.

Apart from the subtraction terms for
double-unresolved configurations, the remaining contributions to the
subtraction counterterms include various soft and collinear limits that
either involve kinematic configurations that are significantly simpler
than the double-unresolved ones or correspond to soft-collinear configurations
which are relatively easy to deal with.

Finally, we note that the subtraction terms and their integrals over
unresolved phase spaces involve \emph{universal} functions that appear in
soft and collinear limits of matrix elements; hence, they are fully
determined by the number, types and
color charges of external particles, and are independent of the matrix elements of hard
processes. Since factorizable corrections to weak boson fusion
process involve a $t$-channel momentum transfer by a colorless vector
boson, they are topologically similar to QCD corrections in deep
inelastic scattering (DIS), as mentioned previously. The calculation of
integrated subtraction terms for DIS has been performed in the
context of the nested soft-collinear subtraction scheme in
Ref.~\cite{Asteriadis:2019dte}. We can then use results obtained in
that reference for the WBF case as well. 
We refer the reader to Ref.~\cite{Asteriadis:2019dte}
for further details regarding the subtraction scheme.

\begin{figure}
\centering
\includegraphics[width=198pt]{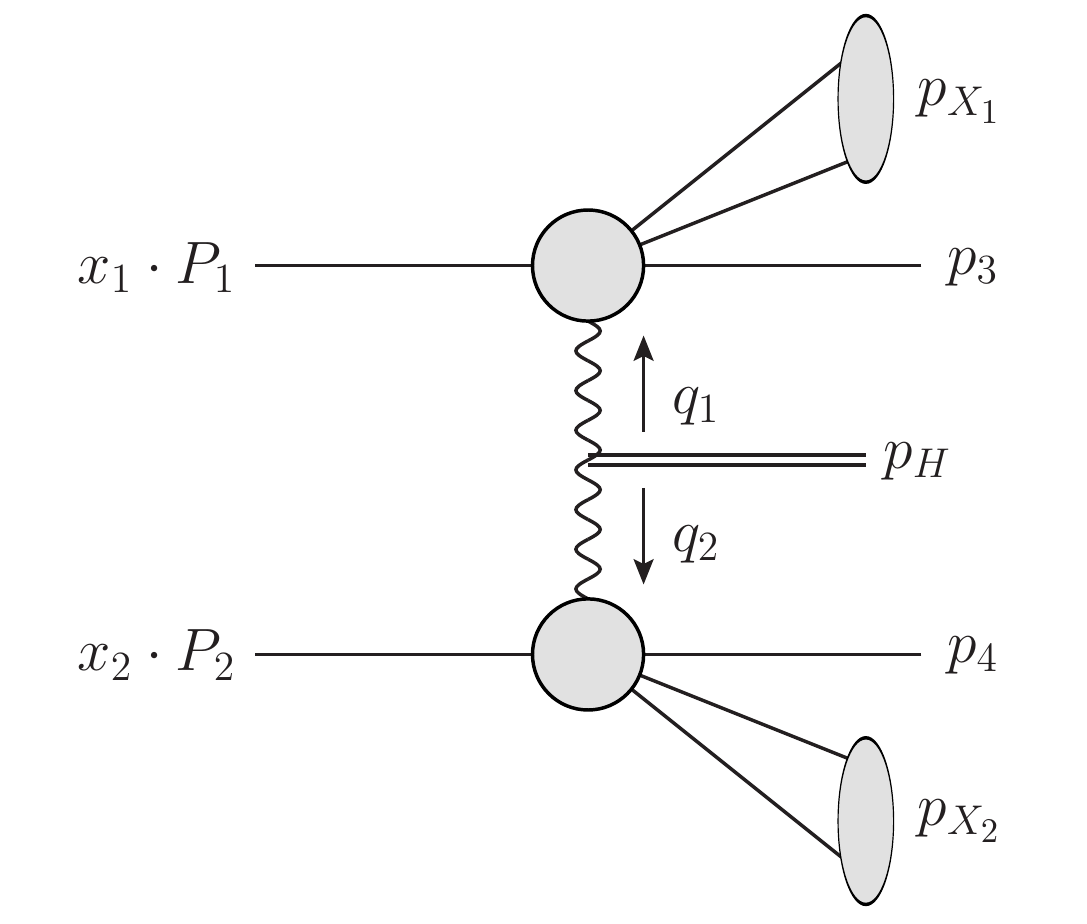}
\caption{Conventions used for the generation of the phase space to
  compute factorizable contributions to  Higgs production in WBF.  Momenta $p_{X_1}$ and $p_{X_2}$
  are combined momenta of all additional (massless) partons radiated from
  the upper or the lower line, respectively}
\label{fig:ps:fact:generic}
\end{figure}

It is well-known that one of the challenging issues pertinent to the
computation of higher order QCD corrections to weak boson fusion is
the phase-space parametrization. Such a parametrization should, on the
one hand, lead to an efficient sampling of the phase space and,
therefore, to an efficient Monte-Carlo integration and, on the other
hand, it should also allow for a seamless connection with the
subtraction scheme. Because of this, we now briefly describe how the
phase space is parametrized in our computation.

At variance with earlier applications of the nested soft-collinear
subtraction scheme, we work in the laboratory frame, i.e. in the 
center-of-mass frame of the incoming \emph{protons}.  We consider the
generic process shown in Fig.~\ref{fig:ps:fact:generic}.  We assume
that particles with momenta $p_3$ and $p_4$ are ``hard'' and particles
collectively denoted with $p_{X_{1,2}}$ can become soft and collinear
to other particles.
The separation into hard and soft particles either happens naturally
(for example, for the NNLO subprocess $q q\to Hqq + gg$ the hard
partons are the two final-state quarks while the partons that can
become soft and collinear are the gluons) or it is achieved by
introducing additional partition functions (as it happens in the case
of several quarks in the final state). An analogous procedure was
carried out for computing the NNLO QCD corrections to DIS
\cite{Asteriadis:2019dte} and we make use of it in the current
calculation.

Using the notation in Fig.~\ref{fig:ps:fact:generic}, we schematically
write the weak boson fusion cross sections in the following way
\begin{align}
\begin{split}
  {\rm \sigma} & = {\cal N}\int_0^1 \frac{\textrm{d}x_1 \textrm{d}x_2 f_1(x_1)
    f_2(x_2)}{2\sh x_1 x_2} \int \dq{3} \dq{4} \dq{X_1} \dq{X_2}
  \dq{H} \\ & \eqspace\times (2\pi)^d \delta^{(d)} \left( P_i - P_f \right)
  \msq{x_1 P_1, x_2 P_2}{p_3, p_{X_1}, p_4,p_{X_2},p_H} \, ,
\end{split}
\label{eq1}
\end{align}
where ${\cal N}$ is the (possible) symmetry and averaging factor,
$P_{1,2}$ are the protons' momenta, $P_i = x_1 P_1 + x_2 P_2$,
$P_f = p_3 + p_4+p_{X_1} + p_{X_2} + p_H$ and
$\sh = (P_1 + P_2)^2$ is the
\emph{ hadronic} center-of-mass energy. 
Also, $f_{1,2}$ are parton
distribution functions, ${\cal M}$ is the relevant matrix element that
depends on the momenta of initial- and final-state particles and
$[{\rm d} p]$ is the phase-space element for a particle or a
collection of particles with momenta $p$. More precisely, we define
\begin{equation}
  \dq{i} = \frac{{\rm d}^{d-1} p_i}{(2\pi)^{d-1} 2 E_i},
\end{equation}
for a  single particle and 
\begin{equation}
  \dq{X_i} = \prod\limits_{i\in X_i}\frac{{\rm d}^{d-1}
    p_i}{(2\pi)^{d-1} 2 E_i},
\end{equation}
for several  particles. Finally, we set $\dq{X_i}=1$ if $X_i$ does not
contain any particle.

To proceed further, we introduce two auxiliary momenta $q_{1,2}$ defined as
follows (see Fig.~\ref{fig:ps:fact:generic})
\begin{align}
\label{eq:ps:fact:gen:qi}
\begin{split}
q_1 = p_3 + p_{X_1} - x_1  P_1 \, , ~~~~~~~
q_2 = p_4 + p_{X_2} - x_2  P_2 \, .
\end{split}
\end{align}

We then insert unity into the phase-space integrand and write 
\begin{align}
\label{eq:ps:nlo:mu}
 1 = \int_0^{\mu^2_\textrm{max}} \textrm{d}\mu_i^2 \ \delta(\mu_i^2 +
 q_i^2) \, ,
\end{align}
for $i = 1,2$.  We note that the relative sign between $\mu_i^2$ and
$q_i^2$ in the argument of the delta-function is due to the fact that
the momenta $q_{1,2}^2$ are space-like. In the above formula $\mu^2_{\rm
  max}$ is an arbitrary quantity with the dimension of energy
squared. We note that $\mu_{\rm max}$ should be large enough to
accommodate all the relevant values of $|q_{1,2}^2|$; we take it to be
$\mu^2_\textrm{max} = s_{
  \rm had}- m_H^2$, where $m_H$ is the mass of the
Higgs boson.

The benefit of having the delta-function in Eq.~(\ref{eq:ps:nlo:mu}) 
to appear in the WBF phase space follows from the fact that its argument
is linear in $x_{i}$, $i=1,2$. Indeed, using explicit expressions for
$q_{1,2}$ and the fact that $P_i^2 = 0$, we find
\begin{align}
\begin{split}
\delta(\mu_i^2 + q_i^2) &= \delta(\mu_i^2 + (p_{3,4} + p_{X_i} - x_i
P_i)^2) \\ &= \delta(\mu_i^2 + (p_{3,4} + p_{X_i})^2 - 2 x_i (p_{3,4}
+ p_{X_i}) P_i) \\ &= \frac{1}{|2 (p_{3,4} + p_{X_i}) P_i|}
\ \delta{\left(\frac{\mu_i^2 + (p_{3,4} + p_{X_i})^2}{2 (p_{3,4} +
    p_{X_i}) P_i} - x_i \right)} \, .
\end{split}
\end{align}

We can then use these expressions to integrate over $x_{1,2}$ in
Eq.~(\ref{eq1}).  We find
\begin{align}
  \begin{split} 
   {\rm \sigma} & = \int \limits_{0}^{\mu_{\rm max}^2} {\rm d} \mu_1^2
   \,{\rm d}\mu_2^2 \int \frac{\dq{3} \dq{4} \dq{X_1} \dq{X_2}
     \dq{H}}{2\sh x_1^* x_2^*} \; \frac{ f_1(x_1^*) f_2(x_2^*)}{|2
     (p_{3} + p_{X_1}) P_1| |2 (p_{4} + p_{X_2}) P_2| } \\ &\eqspace\times
   (2\pi)^d \delta^{(d)}\left( P_i^*- P_f \right ) 
   \msq{x_1^* P_1, x_2^* P_2}{ p_3, p_{X_1}, p_4,p_{X_2},p_H} \, ,
\end{split} 
\label{eq5}
\end{align}
where $P_i^* = x_1^* P_1 + x_2^* P_2$ and 
\begin{equation}
x_i^* = \frac{\mu_i^2 + (p_{3,4} + p_{X_i})^2}{2 (p_{3,4} + p_{X_i})
  P_i}, \;\;\;\;\;\;\; i=1,2.
\label{eq6}
\end{equation}
We require, of course, that $x_i^* \in [ 0,1]$; if this is not the
case, the corresponding kinematic points are rejected.

We remove the delta-function in Eq.~(\ref{eq5}) by integrating over the
four-momentum $p_H$ of the Higgs boson. We use
\begin{align}
  \begin{split}
&\dq{H} \ (2\pi)^d \delta^{(d)}{\left(x_1 P_1 + x_2 P_2 - p_3 - p_4 -
      p_{X_1} - p_{X_2} - p_H\right)} \\ &\eqspace= 2\pi \, \delta\big((q_1 +
    q_2)^2 - m_H^2\big) 
    = 2\pi \, \delta\big(-\mu_1^2 - \mu_2^2 + 2 (q_1 q_2)
    - m_H^2\big) \, ,
\end{split}
\end{align}
and remove this last delta-function by integrating over $\mu_2^2$.  We
should, however, account for the fact that $q_2$ depends on $x_2^*$
which, in turn, depends on $\mu_2^2$ implicitly.

Integrating over $\mu_2^2$, we find 
\begin{align} 
\begin{split} 
  \sigma &= \int \limits_0^{\mu_\textrm{max}^2} \textrm{d}\mu_1^2
  \int \dq{3} \dq{4} \dq{X_1} \dq{X_2} \;
  \frac{ f_1(x_1^*) f_2(x_2^*) }{2\sh x_1^*x_2^*} \times  \frac{2\pi }{|4 (q_1 P_1)(p_H P_2)|} \\
  &\eqspace\times
  \msq{x_1^* P_1, x_2^* P_2}{ p_{3}, p_4,\dots} ,
\end{split}
\label{eq8}
\end{align}
  where $p_H=q_1+q_2$ and 
\begin{align}
\begin{split}
\mu_2^2 = \frac{(p_4 + p_{X_2})  P_2}{ (p_4 + p_{X_2}+q_1)
     P_2} \times \left[-m_H^2 -\mu_1^2 + \bigg( 2(p_4
  + p_{X_2})q_1 - \frac{ (p_4 + p_{X_2})^2 (q_1 P_2) }{ (p_4 +
    p_{X_2})  P_2} \bigg)\right] \, 
\label{eq9}
\end{split}
\end{align}
should be used when computing $x_2^*$ and $q_2$. 

The expression for the cross section in Eq.~(\ref{eq8}) is general; it
can be used at any order in perturbation theory and for various
combinations of emissions off initial and final states.  For our
calculation, we will mainly need to consider the case of two 
emissions off the upper (lower)  quark line and, consequently, no emissions off
the lower (upper) quark line.  Focusing on the emissions off the upper line,  we set
$\dq{X_2} \to 1$ in the expression for the phase space in
Eq.~(\ref{eq8}), and $p_{X_2} \to 0$ in Eq.~(\ref{eq9}).  Since $p_4^2
= 0$, the expression for $\mu_2$ simplifies to
\begin{equation}
\mu_2^2 = \frac{ (p_4 P_2)}{ (p_H P_2)} \times \left[ m_H^2 +
  \mu_1^2 - 2 (q_1 p_4) \right].
\label{eq12a}
\end{equation}
Finally, we note that if at NLO or at NNLO additional emissions occur
only at the lower line in Fig.~\ref{fig:ps:fact:generic}, it is of
course beneficial to integrate over $\mu_1^2$, instead of
$\mu_2^2$. We do not write an explicit formula for the cross section
in this case as it can simply be obtained by a trivial re-labeling of
various particles in the expressions given in
Eqs.~(\ref{eq8},\ref{eq9}).

Having derived a suitable parametrization of the weak boson fusion
phase space, we will have to use it to construct both the subtraction
terms and their integrated counterparts required to extract and
regulate all the relevant infra-red and collinear divergences.  To
this end, we would like to use the results for DIS that we have derived in
Ref.~\cite{Asteriadis:2019dte}.  However, in that reference the
subtraction framework was constructed in the \emph{partonic}
center-of-mass frame. Since the subtraction formalism of
Ref.~\cite{Caola:2017dug} is not manifestly boost invariant, it is not
immediately obvious that one could use the results of
Ref.~\cite{Asteriadis:2019dte} here without modifications.

To illustrate the connection between the phase space parametrization in
Eq.~(\ref{eq8}) and Ref.~\cite{Asteriadis:2019dte}, we describe
how the construction of a typical subtraction term derived in
Ref.~\cite{Asteriadis:2019dte} would proceed.  For the sake of
argument, we consider the emission of two gluons with momenta $p_{5,6}$
off the upper (quark) line in Fig.~\ref{fig:ps:fact:generic} and study
the limit when $p_6$ is collinear to the incoming proton $P_1$.  The
operator that extracts the relevant limit from the cross section is
referred to as $C_{61}$ in Ref.~\cite{Asteriadis:2019dte}. In that reference,
we worked at fixed $x_1,x_2$ and wrote the collinear subtraction term as
\begin{equation}
\begin{split}
  C_{61} \left[ \sigma \right ] & = g_s^2\int \frac{{\rm d} x_1 
    {\rm d} x_2 f(x_1) f(x_2)}{2\sh x_1 x_2} \; \int
  \prod\limits_{i=3}^6\dq{i} (2\pi)^d \delta^{(d)} \left ( z x_1 P_1
  +x_2 P_2-p_3-p_4-p_5 \right ) \\ &  \eqspace \times \frac{1}{(P_1 p_6) } \; P_{qq}
  \left (z \right ) \; \frac{ \msq{z x_1 P_1, x_2 P_2}{p_3,p_4,p_5}}{z x_1} \, ,
\end{split}
\label{eq11}
\end{equation}
where $g_s$ is the (bare) strong coupling constant and $P_{qq}$ is the standard
($d$-dimensional) LO splitting function whose precise form is
irrelevant for our discussion. The variable $z$ is related to the
energy of the gluon with momentum $p_6$ as $E_6 = x_1 E_1(1-z)$, where
$E_1$ is the energy of the incoming \emph{proton}. 

We then write
\begin{equation}
  \dq{6} = \frac{{\rm d} \Omega_6}{2(2\pi)^{d-1}} \;
  (x_1 E_1)^{2-2\ep} {\rm d} z (1-z)^{1-2\ep},
\end{equation}
and use this expression in Eq.~(\ref{eq11}) to find (see
Ref.~\cite{Asteriadis:2019dte} for details)
\begin{equation}
\begin{split}
  C_{61} \left[ \sigma \right ] &= -[\alpha_s]
  \frac{\Gamma(1-\ep)^2}{\Gamma(1-2\ep)\ep} \int \frac{{\rm d} x_1
    {\rm d} x_2 f(x_1) f(x_2)}{2\sh x_1 x_2}\; (2 x_1 E_1)^{-2\ep} \;
  \int {\rm d} z \; (1-z)^{-2\ep} \; P_{qq} \left (z \right ) \; \\
       &\eqspace\times\int \prod_{i=3}^5 \dq{i} \; (2\pi)^d \delta^{(d)} \left ( z x_1 P_1 +x_2
       P_2-p_3-p_4-p_5 \right ) \\
       &\eqspace\times\frac{ \msq{z x_1 P_1, x_2 P_2}{p_3,p_4,p_5} }{z} \, ,
\end{split}
\label{eq13}
\end{equation}
where  we have defined
$[\as]=({g_s^2}/{8\pi^2})\times{(4\pi)^\ep}/{\Gamma(1-\ep)}$.

We now repeat the same computation using the phase-space
parametrization in Eq.~(\ref{eq8}).  The difference with what we just
did originates from the fact that in  Eq.~(\ref{eq8}) the Bjorken
variables $x_{1,2}$ are not free parameters anymore but, rather, are
determined from the kinematics of final state particles.  Hence, when
we extract the collinear limits of the cross sections, we have to act
on $x_{1,2}$ as well.

Using  Eq.~(\ref{eq6}), we find 
\begin{equation}
  C_{61} \left [ x_1^* \right ] = {\bar x_1}^*= x_1^{\rm NLO} +
  \frac{E_6}{E_1},
\end{equation}
where  
\begin{equation}
  x_1^{\rm NLO} = \frac{\mu_1^2 + (p_3+p_{5})^2}{2(p_3+p_{5} ) P_1},
\end{equation}
is the Bjorken $x$ that one would have obtained by performing a NLO
calculation with a parton $p_5$ in the final state.  We also find that
\begin{equation}
C_{61} \left [ x_2^* \right ] =\frac{ C_{61} \left [ \mu_2^2 \right ]
}{2 (p_4 P_2)} \equiv x_2^{\rm NLO},
\end{equation}
and that the collinear limit of $\mu_2^2$ is determined by the
collinear limit of $q_1$ which is given by
\begin{equation}
q_1 \to {\bar q}_1 = p_3 + p_{X_1} - {\bar x}_1^* P_1 = p_3 + p_{5}
- x_1^{\rm NLO} P_1.
\end{equation}
Again, this is the expression for the momentum transfer that one
obtains in the NLO computation where an additional parton with momentum
$p_5$ is emitted.

We can now compute the collinear limit of the cross section using the
phase-space parametrization Eq.~(\ref{eq8}). We find 
\begin{equation}
\begin{split} 
   C_{61} \left [ {\rm d} \sigma \right ] & = g_s^2\int {\rm d} \mu^2
   \int \dq{3}\dq{4}\dq{5}\dq{6} \; \frac{f_1({\bar x_1}^* )\;
     f_2({\bar x_2}^*) }{2\sh {\bar x_1}^* {\bar x_2}^* }
   \frac{2\pi}{|4 (\bar q_1 P_1) (p_H P_2)|}
   \\
   &\eqspace\times\delta_{\rm mom-cons}\times
   \frac{1}{ {\bar x_1}^* (P_1 p_6) } \times
       \frac{{\bar x_1}^* E_1 }{
         {\bar x_1}^* E_1 - E_6 } P_{qq}{\left( \frac{
       {\bar x_1}^* E_1 - E_6}{{\bar x_1}^* E_1} \right)} \\
   &\eqspace \times \msq{\left[{\bar x_1}^* - E_6/E_1\right] P_1,{\bar x}_2^* P_2}{\dots, p_5} \, ,
\label{eq18}
\end{split}
\end{equation}
where we have used $P_{qq}(1/z) = -P_{qq}(z)/z$ and defined
\begin{equation}
  \delta_{\rm mom-cons} = (2\pi)^d\delta^{(d)}\big( \left[
        {\bar x_1}^*- E_6/E_1 \right] P_1 + {\bar x}_2^* P_2 - p_3-p_4-p_5\big) \, .
        \label{eq:momcons}
\end{equation}
The reduced matrix element squared simplifies 
\begin{align} 
  \begin{split}
  	\msq{\left[{\bar x_1}^* - E_6/E_1\right] P_1,{\bar x}_2^* P_2}{\dots, p_5} =
  	\msq{x_1^{\rm NLO} P_1, x_2^{\rm NLO}
  		P_2}{\dots,p_5} \, .
\end{split}
\end{align}
An analogous simplification occurs in the argument of the 
delta-function in Eq.~\eqref{eq:momcons} where we find 
\begin{equation}
\delta_{\rm mom-cons} = (2\pi)^d\delta^{(d)}\big(
        x_1^{\rm NLO} P_1 + x_2^{\rm NLO} P_2 - p_3-p_4-p_5\big).
\end{equation}

To simplify Eq.~(\ref{eq18}) further, we introduce a new variable $z$
\begin{equation}
z = \frac{ {\bar x_1}^* E_1 - E_6}{{\bar x_1}^* E_1} = \frac{x_1^{\rm
    NLO}}{x_{1}^{\rm NLO} + \frac{E_6}{E_1} },
\end{equation}
and express the energy of the collinear  gluon through $z$. We find 
\begin{equation}
E_6 = E_1 \;  x_1^{\rm NLO} \; \left(\frac{1-z}{z}\right),
\end{equation}
and ${\bar x_1}^* = {x_1^{\rm NLO}}/{z}$.

As a result, the collinear limit of the cross section
becomes 
\begin{equation}
\begin{split} 
  C_{61} \left [ {\rm d} \sigma \right ] & = g_s^2 \int {\rm d} \mu^2
  \int \prod_{i=3}^6 \dq{i} \; \frac{f_1(x_1^{\rm NLO}/z ) \;
    f_2(x_2^{\rm NLO} ) }{2\sh x_1^{\rm NLO} x_2^{\rm NLO}}
  \frac{z^2}{ x_1^{\rm NLO} (P_1 p_6) }\; \\ 
  & \eqspace \times \delta_{\rm mom-cons} \times
  \frac{P_{qq}(z)}{z}
  \msq{x_1^{\rm NLO} P_1 ,x^{\rm NLO}_2 P_2}{\dots,p_5} \\
  &\eqspace\times \frac{2\pi}{|4 (\bar q_1 P_1) (p_H  P_2)|}.
\label{eq23a}
\end{split}
\end{equation}
Writing
\begin{equation}
\dq{6} = \frac{{\rm d} \Omega_{6}}{2(2\pi)^{d-1}} E_6^{1-2\ep} {\rm d}
E_6 = \frac{{\rm d} \Omega_{6}}{2(2\pi)^{d-1}} \; \left [ \frac{E_1
    x_1^{\rm NLO} (1-z)}{z} \right ]^{1-2\ep} \frac{E_1 x_1^{\rm NLO}
  {\rm d} z}{z^2},
\end{equation}
and using this expression in Eq.~(\ref{eq23a}), we arrive at
\begin{equation}
\begin{split}
  C_{61} \left[{\rm d} \sigma\right] & = -[\alpha_s]
  \frac{\Gamma(1-\ep)^2}{\Gamma(1-2\ep)\ep} \int {\rm d} \mu^2 \int
  \prod\limits_{i=3}^{5}\dq{i} \; \int {\rm d} z (1-z)^{-2\ep} \left ( \frac{
    2 E_1 x_1^{\rm NLO}}{z} \right )^{-2\ep} \\ &
  \eqspace\times\delta_{\rm mom-cons} \times
  \frac{P_{qq}(z)}{z}\frac{ \msq{x_1^{\rm NLO} P_1 ,x^{\rm NLO}_2 P_2}{\dots,p_5}}{2\sh{x_1}^{\rm NLO} x_2^{\rm NLO} } \\
&\eqspace\times
  f_1(x_1^{\rm NLO}/z ) \; f_2(x_2^{\rm NLO} ) \frac{2\pi}{|4 (\bar q_1 P_1) (p_H P_2)|}.
\end{split}
\label{eq25}
\end{equation}

The above equation and Eq.~(\ref{eq13}) look different. However it is easy to see that by
using the phase-space parametrization of Eq.~\eqref{eq8} in 
Eq.~\eqref{eq13} and integrating over the Bjorken variables $x_{1,2}$,   one obtains 
Eq.~\eqref{eq25}.

This quick derivation shows that for the kinematic situation when one
of the final state gluons is emitted along the direction of an
incoming quark, one can indeed combine the subtraction terms and the
integrated subtraction terms computed in
Ref.~\cite{Asteriadis:2019dte} with the phase-space parametrization
Eq.~(\ref{eq8}). We have checked that the same holds for all other
limits that are relevant for the NLO and NNLO computations. Apart from 
a more flexible parametrization, this result also provides a
non-trivial check on the robustness of the subtraction procedure of
Ref.~\cite{Caola:2017dug} and allows us to use the subtraction terms derived in Ref.~\cite{Asteriadis:2019dte} without modification.

Finally, an important benefit of the phase-space parametrization described in this section is that both fully-resolved
matrix elements and subtraction counterterms are always calculated in
the same frame, without the need for longitudinal boosts. We find that
this significantly improves the efficiency of the Monte-Carlo
integration of the most complicated subtraction counterterms.
  
\section{NNLO corrections to Higgs production in WBF}
\label{sect3}
In this section we present results for fiducial cross sections
and kinematic distributions for Higgs production in WBF. We begin in Section \ref{sect3a} 
by treating  the Higgs boson as a stable
particle. This is important both for  validating our results against previous
calculations~\cite{Cacciari:2015jma,Cruz-Martinez:2018rod} and as a
reference for the case of the decaying Higgs boson which  we study later.  In the following
subsections, we consider two phenomenologically important Higgs decay
modes, namely $H \to b\bar{b}$ in Section \ref{sect3b} and $H \to WW^* \to \ell^- \bar
\nu \ell^+ \nu$  in Section \ref{sect3c}, and explore the extent to which
additional kinematic cuts applied to  the decay products of the Higgs
boson modify the NNLO QCD corrections.

For all  phenomenological results reported in this paper, we use a baseline setup which is
very similar (though not identical) to the one of
Ref.~\cite{Cacciari:2015jma}. We set the Higgs boson mass to $M_H =
125~\gev$, its width to $\Gamma_H=4.165~\mev$, the vector boson masses to $ M_W = 80.398~\gev$ and $M_Z =
91.1876~\gev$, and their widths to $\Gamma_W = 2.105~\gev$ and
$\Gamma_Z = 2.4952~\gev$. We use the Fermi constant $G_F = 1.16639
\times 10^{-5}~\gev^{-2}$, and approximate the CKM matrix by an
identity matrix. We employ the {\sf NNPDF31-nnlo-as-118} parton
distribution functions~\cite{NNPDF:2017mvq} and use them for all 
calculations reported in this paper, irrespective of the perturbative
order.  We also use the value of the
$\overline {\rm MS}$ strong coupling constant $\alpha_s(M_Z) = 0.118$
as provided by the PDF set. The evolution of parton distribution
functions and the strong coupling constant is obtained from
LHAPDF~\cite{Buckley:2014ana}.  We employ dynamical renormalization and
factorization scales; their central values are set to~\cite{Cacciari:2015jma}
\begin{equation}
\mu_0= \sqrt{ \frac{m_H}{2} \sqrt{\frac{m_H^2}{4} + p_{\perp,H}^2}}.
\label{eq21}
\end{equation}

We consider proton-proton collisions with center-of-mass energy of
13 TeV, and define the WBF fiducial cross section in the following
way.  We reconstruct jets using the inclusive anti-$k_\perp$
algorithm~\cite{Cacciari:2008gp} with $R=0.4$. We then consider the
list $J=\{j_i\}$ of jets satisfying $p_\perp \ge 25~\gev$, $|y|\le
4.5$, and impose the following constraints. For each event, we require
that $J$ contains at least two jets. Furthermore, the two
leading-$p_\perp$ jets in $J$ should be separated by a large
rapidity interval $|y_{j_1} - y_{j_2}| \ge 4.5$,  and their invariant
mass should be larger than $600~{\rm GeV}$. They should also
be in different hemispheres in the laboratory frame;
to enforce this, we require that the product of their rapidities is
negative, $y_{j_1} y_{j_2} \le 0$. 

Before presenting our results, we briefly discuss the checks that we have
performed on our calculation. For stable  Higgs, we have first 
checked the inclusive cross section against a customized version of
the \texttt{proVBFH-inclusive} code~\cite{proVBF}, which is based on
the calculations of Refs.~\cite{Cacciari:2015jma,Dreyer:2016oyx} and
on Ref.~\cite{Salam:2008qg}. We have modified the original program to
be able to compare  individual contributions  separately
and we have found agreement with the corresponding results obtained
from our calculation. 
Moving to fiducial results, we note
that although our setup is very similar to the one in
Ref.~\cite{Cacciari:2015jma}, it is not identical. In particular, we use
updated parton distribution functions and a slightly different jet
selection procedure.  Because of this, the results
presented here cannot be directly compared with those of Ref.~\cite{Cacciari:2015jma}. We have, however, produced
a set of results with the setup of Ref.~\cite{Cacciari:2015jma}, and
found agreement with the fiducial cross-sections reported there at the
level of few per-mille or better. We have also compared differential
distributions, and also in this case found good agreement within the
numerical precision of the calculations. For the $H\to b\bar b$ and
$H\to WW^* \to 4l$ results, we have compared our implementation against
\texttt{MCFM} version
10~\cite{mcfm,Campbell:1999ah,Campbell:2015qma,Campbell:2019dru}
at NLO and LO, respectively, and
found perfect agreement.

\subsection{Stable Higgs boson}
\label{sect3a}

In this section we present the fiducial cross sections for the
production of a stable Higgs boson in WBF at NNLO in QCD.
Analogous results have been already
presented in
Refs.~\cite{Cacciari:2015jma,Cruz-Martinez:2018rod}. There are two
reasons for us to repeat these computations.  First, it allows us to validate 
 our calculation. To this end, as we have mentioned already, we have
performed an extensive comparison with the results of
Ref.~\cite{Cacciari:2015jma} for both inclusive and fiducial cross
sections, and found agreement at the per-mille level or better in all
cases.  We also find excellent agreement with higher-order corrections
to kinematic distributions presented in that reference.
A second reason to consider WBF production with a stable Higgs boson
is that it provides a benchmark against which to  compare the
results obtained by  considering  the decays of the Higgs boson that are discussed  later. 

\begin{figure}
  \centering
  \includegraphics[width=0.328 \textwidth]{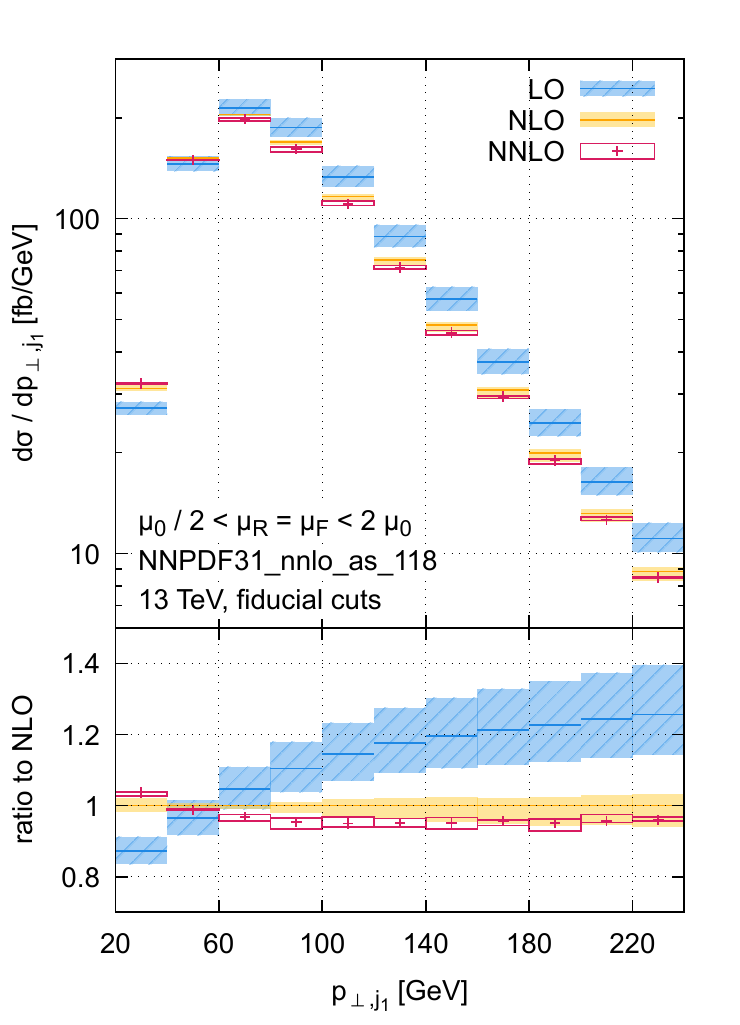}
  \includegraphics[width=0.328 \textwidth]{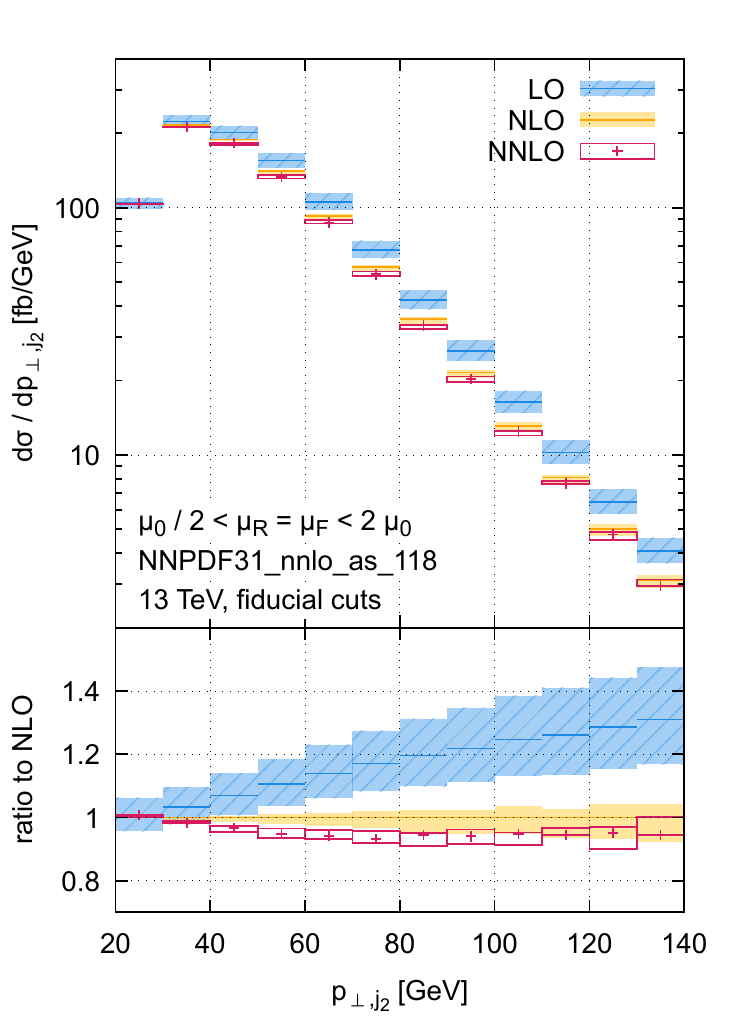}
  \includegraphics[width=0.328 \textwidth]{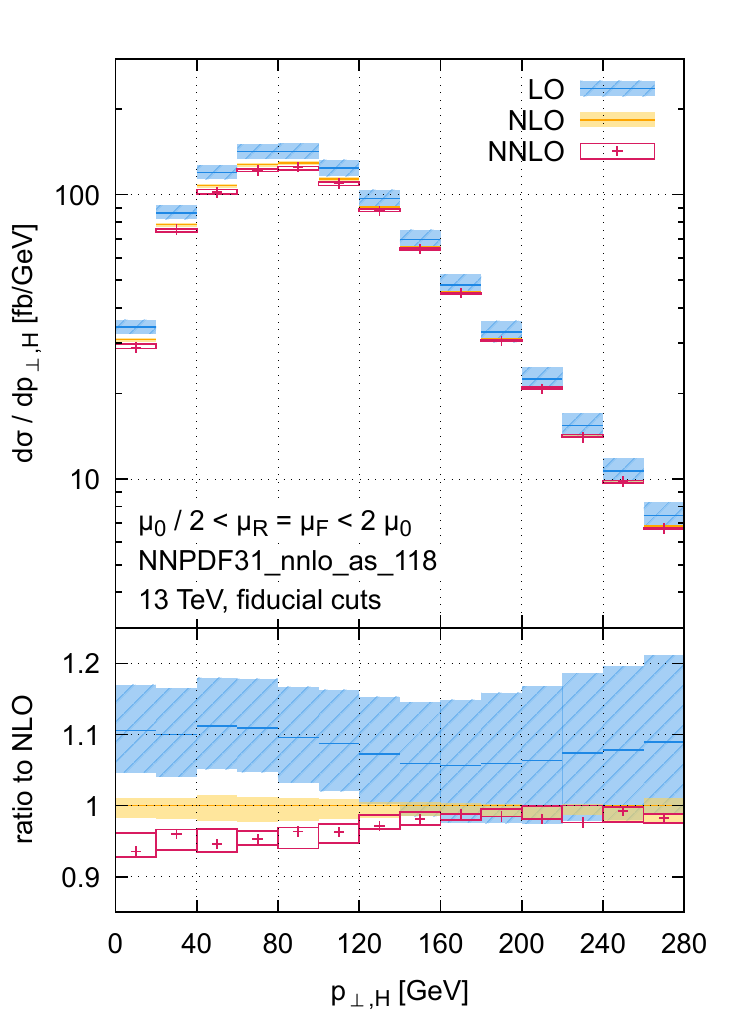}
  \includegraphics[width=0.328 \textwidth]{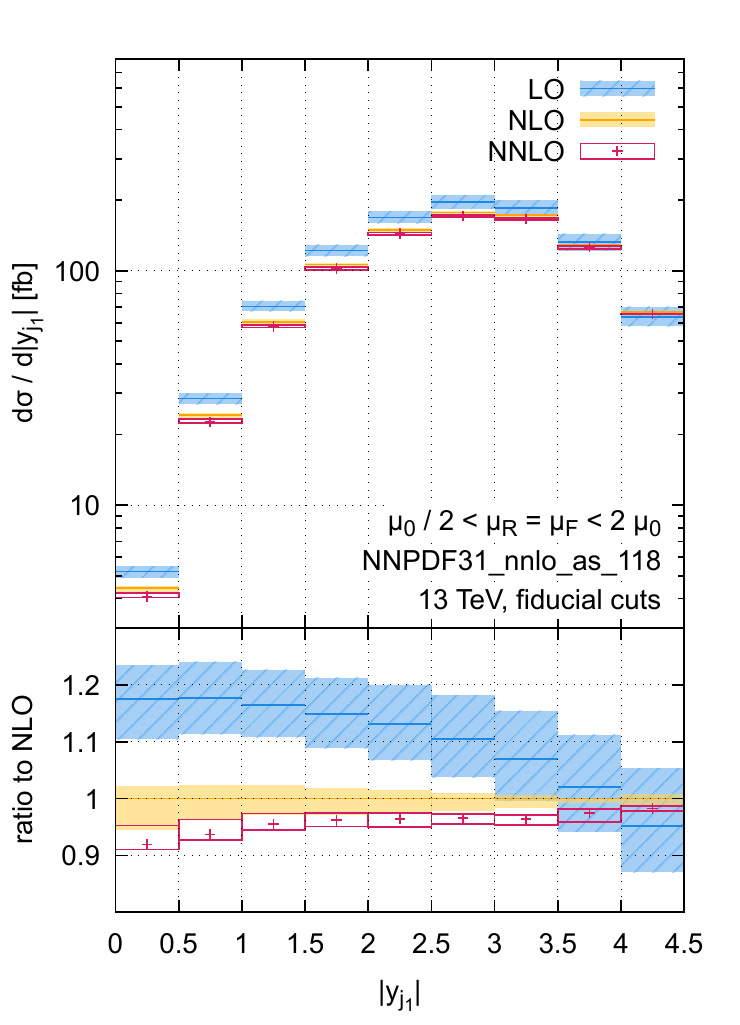}
  \includegraphics[width=0.328 \textwidth]{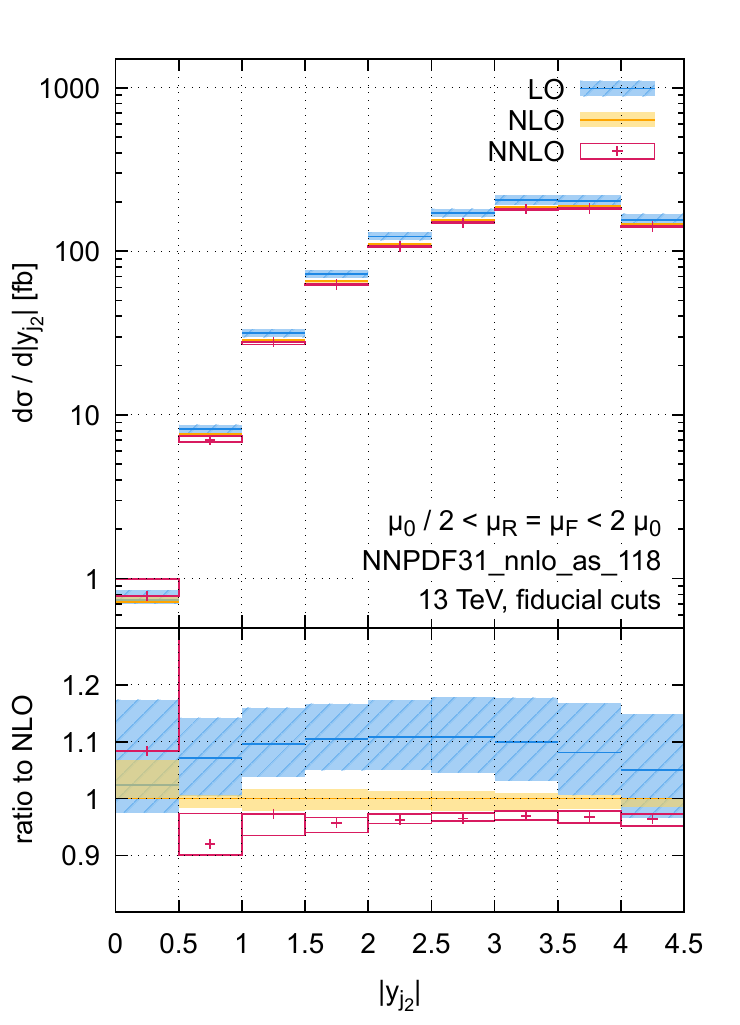}
  \includegraphics[width=0.328 \textwidth]{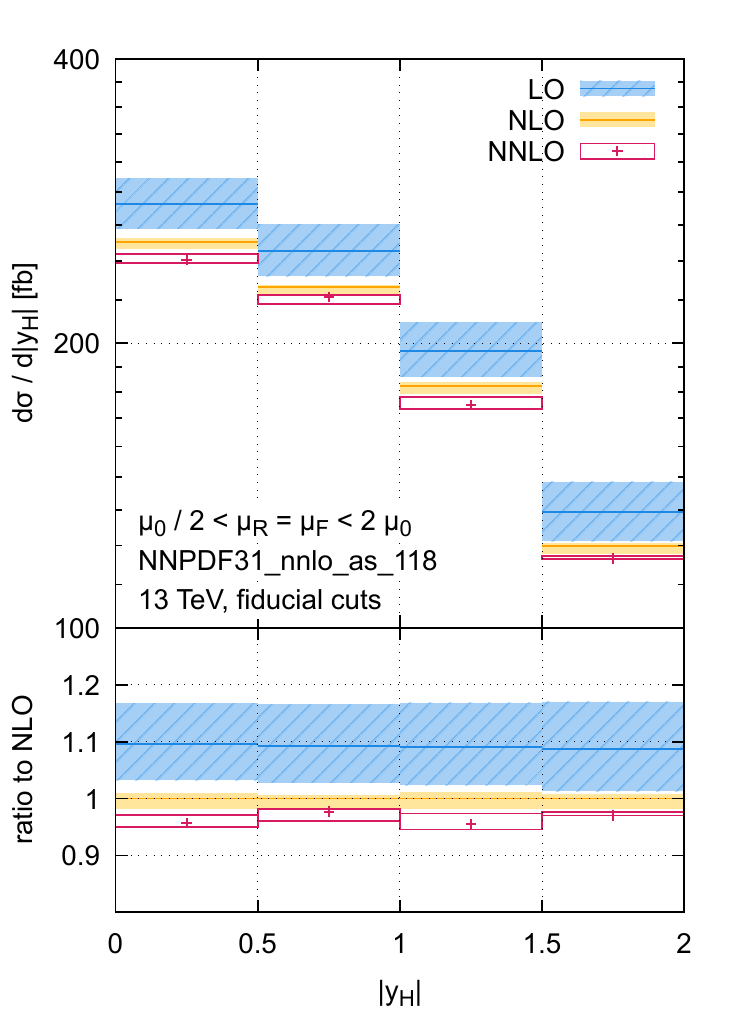}
  \caption{Transverse momenta (upper row) and rapidity (lower row)
    distributions for the leading (left) and subleading (middle) jets
    as well as the Higgs boson (right).
    For each plot, the upper pane displays the LO (hashed blue),
    NLO (solid yellow) and NNLO (red boxes) QCD
    predictions. The lower pane shows the ratio with respect to the NLO
    result at central scale. 
    The lines indicate the central renormalization and factorization
    scale choice, and the bands
    indicate the envelope of the results at different scales.
See text for details.
}
  \label{fig::ptrap}
\end{figure}  

We first report fiducial cross sections at various orders of QCD perturbation theory. Using the setup described above, we obtain
\begin{equation}
\sigma_{\rm LO} = 971^{-61}_{+69}~{\rm fb}, \;\;\;\; \sigma_{\rm NLO}
= 890^{+8}_{-18}~{\rm fb}, \;\; \;\;\;\;\; \sigma_{\rm NNLO} =
859^{+8}_{-10}~{\rm fb}.
\label{eq22}
\end{equation}
The central values of fiducial cross sections are computed with the
renormalization and factorization scales of Eq.~(\ref{eq21}) and
uncertainties are obtained by varying this scale by a factor of two in
both directions.  In Eq.~(\ref{eq22}) and throughout this paper, the
sub- and super-scripts indicate the results computed with $\mu =
\mu_0/2$ and $\mu = 2 \mu_0$, respectively.

Results for cross sections show a familiar
pattern~\cite{Cacciari:2015jma,Cruz-Martinez:2018rod} -- NLO and NNLO
QCD corrections to fiducial cross sections are relatively small; they
decrease the previous order cross section by about 8 and 3.5 percent,
respectively.  The uncertainty of $\sigma_{\rm NNLO}$ as estimated
from scale variation is about a percent, only marginally smaller
than the NLO one. Although we do not show inclusive results here, we
note that NNLO corrections to the inclusive WBF cross section are about one percent and, therefore,
are significantly 
smaller than corrections to the fiducial cross section. This is not surprising, since WBF fiducial
cuts induce a  sensitivity of theoretical predictions to  the non-trivial jet dynamics present in 
this process, see e.g. the discussion in Ref.~\cite{Rauch:2017cfu}.

We now present results for kinematic distributions.  In
Fig.~\ref{fig::ptrap} NNLO QCD results for the transverse momentum and
the rapidity distributions of the leading and subleading jets, as well
as the Higgs boson, are shown.  The upper panes in plots in
Fig.~\ref{fig::ptrap} display LO, NLO and NNLO QCD predictions in hashed
blue, yellow, and red boxes, respectively.  The lines indicate the
central renormalization and factorization scale choice $\mu = \mu_0$
in Eq.~(\ref{eq21}), and the bands indicate the envelope of the results at scales $\mu\in \{\mu_0/2,\mu_0,2\mu_0\}$.  For transverse momenta distributions,
we observe that in all cases the NNLO/NLO bin-by-bin $K$-factors are
reasonably flat, at variance to the NLO/LO ones that exhibit strong dependencies
on transverse momenta of the leading and subleading jets. For the Higgs rapidity distribution, both NLO/LO and NNLO/NLO
$K$-factors are approximately flat in the bulk of the distribution and
consistent with the corrections to fiducial cross sections reported in
Eq.~(\ref{eq22}). For the jets rapidity distributions, the
NLO/LO K-factors have a non-trivial shape (especially in the leading jet case) but the NNLO/NLO ones are fairly
flat. 

\begin{figure}[t]
  \centering
  \includegraphics[width=0.328 \textwidth]{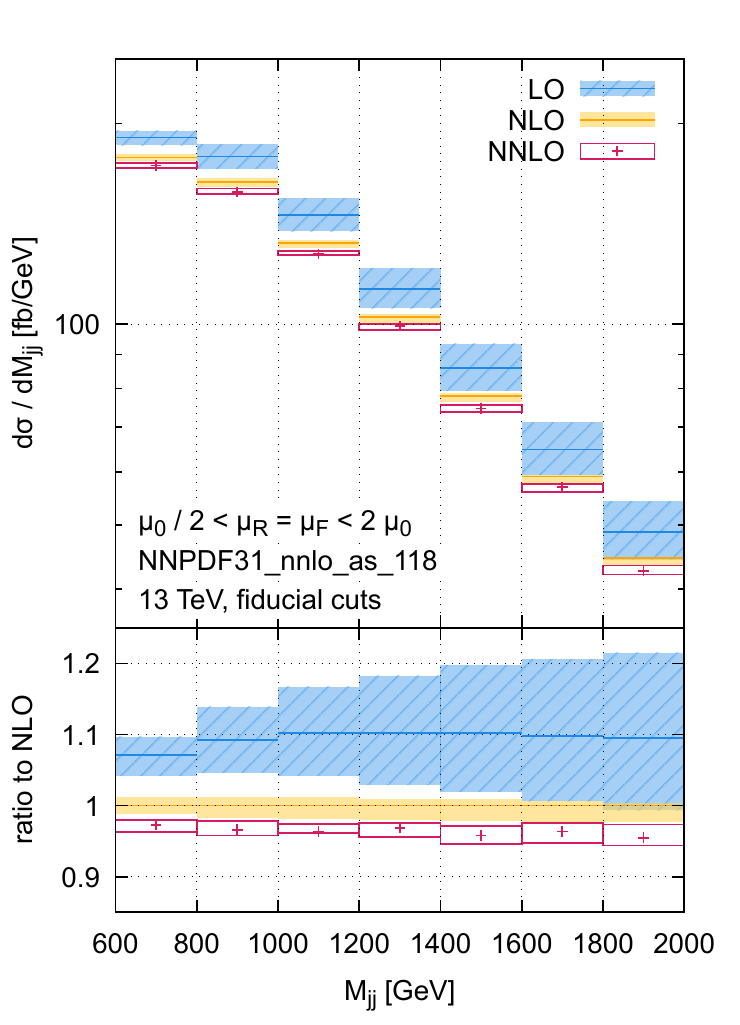}
  \includegraphics[width=0.328 \textwidth]{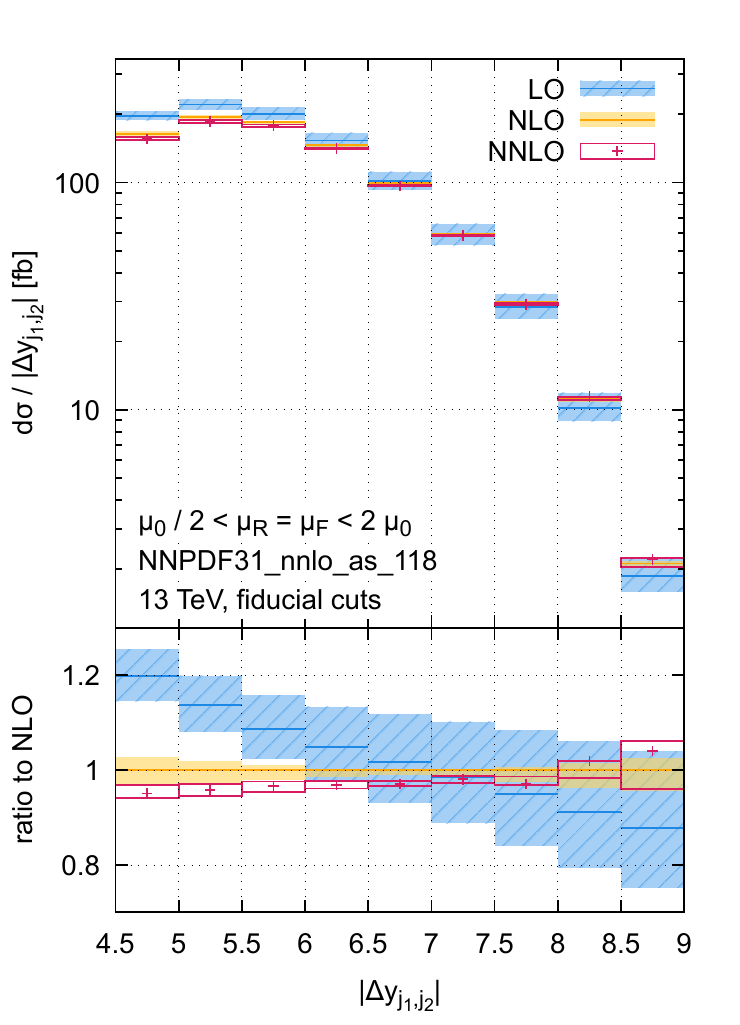} \\
  \includegraphics[width=0.328 \textwidth]{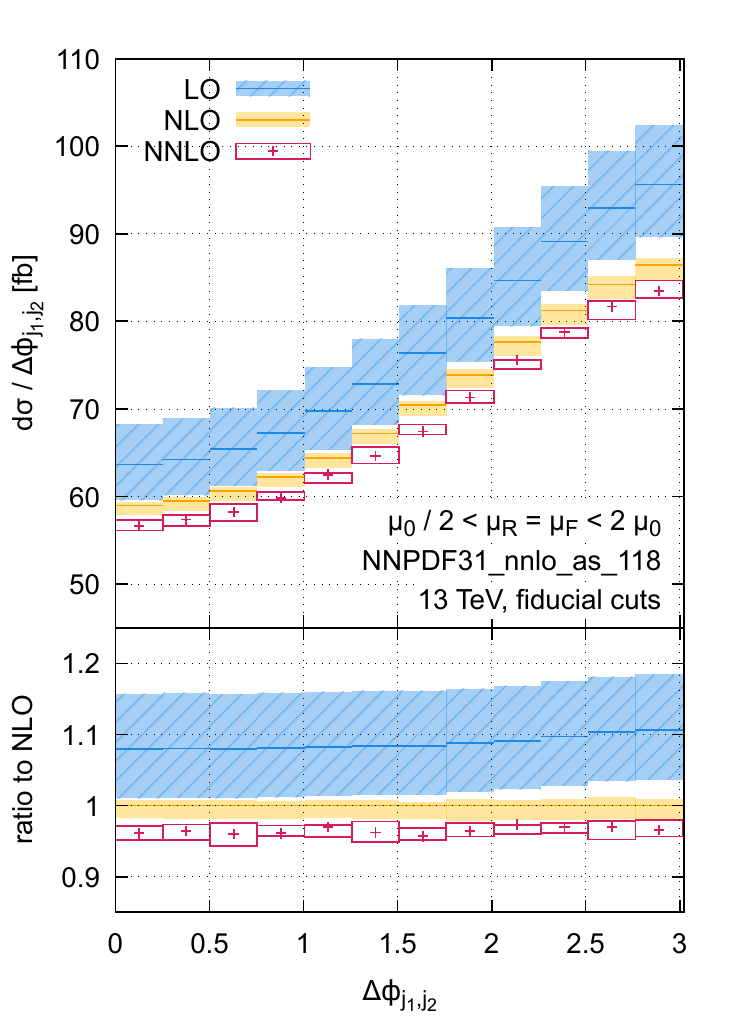}
  \includegraphics[width=0.328 \textwidth]{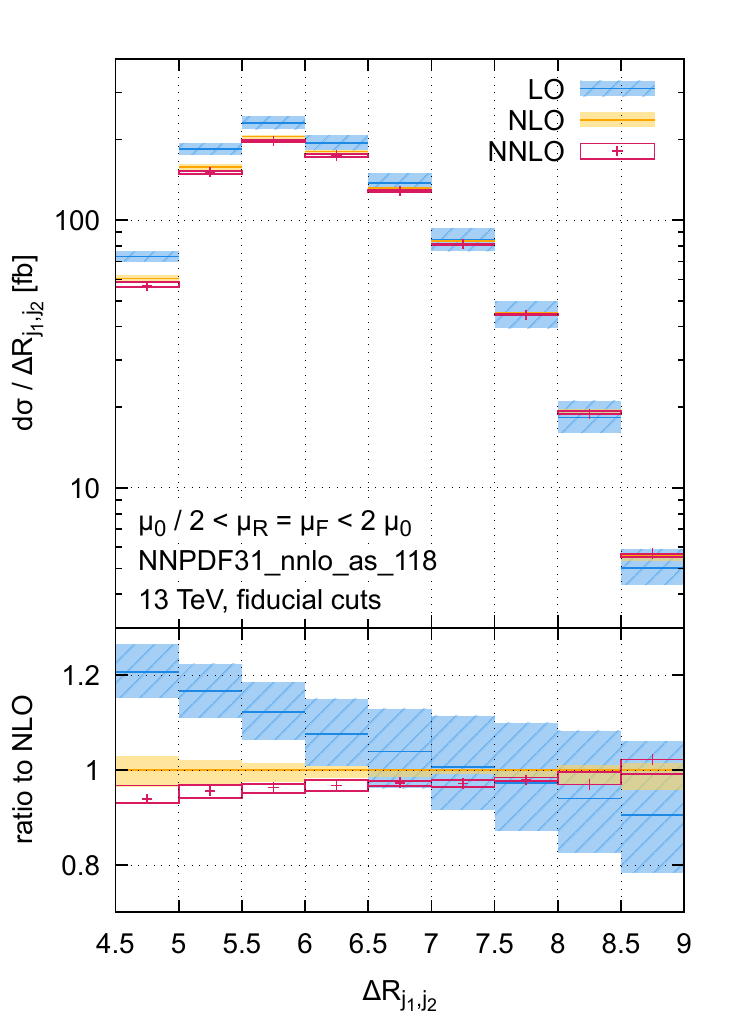} \\
\caption{Kinematic distributions of observables that involve the two
  leading jets for a stable Higgs. Top row: dijet invariant mass
  distribution (left) and rapidity separation (right).
  Bottom row: azimuthal (left) and $\Delta R$ (right) separations.
  See text for details.}
\label{fig::jet12}
\end{figure}

The identification of the weak boson fusion process exploits kinematic
features of the two leading jets.  It is therefore interesting to
check how kinematic distributions that involve them get modified in higher orders
of perturbation theory. In Fig.~\ref{fig::jet12} we display the
invariant mass distribution of the two leading jets, their rapidity
difference $\Delta y_{j_1 j_2} = |y_{j_1} - y_{j_2}|$ and the relative
azimuthal angle $|\phi_{j_1} - \phi_{j_2}|$ distributions, as
well as the distribution of the distance between the two jets in the
transverse plane $\Delta R_{j_1, j_2} = \sqrt{ (y_{j_1} - y_{j_2})^2 +
  (\phi_{j_1} - \phi_{j_2})^2}$.  The distribution of the dijet
invariant mass and the $\Delta \phi_{j_1j_2}$ distribution exhibit
rather flat $K$-factors, both at NLO and at NNLO. By contrast, the
NLO/LO $K$-factors for the rapidity difference and $\Delta
R_{j_1 j_2}$ show stronger dependencies on the corresponding variables,
but they stabilize at NNLO.

In summary, we find that although NLO QCD corrections can have an important
impact on the shapes of differential distributions of Higgs production in WBF, the impact of NNLO corrections is much
milder and it is captured by an overall K-factor to a reasonable approximation.
However, the NNLO distributions often lie outside of the NLO scale variation bands, although
distances  between the NLO and NNLO  bands are small.
We note that  this is also the case for the fiducial cross sections
shown in Eq.~\eqref{eq22}.

\subsection{Results for $H \to b \bar b$}
\label{sect3b}
In this section we study NNLO QCD corrections to Higgs production in
WBF, taking into account Higgs boson decays to a $b\bar{b}$-pair. The
radiative corrections acquire a dependence on the kinematics of Higgs
boson decay products because their four-momenta are used to define the 
selection criteria, as we now explain.

In case of $H \to b \bar b$ decays, the Higgs boson is identified and
reconstructed through an observation of two $b$-jets. Since the $b$-quarks
from Higgs boson decays sometimes get clustered with other final
state partons and since the Higgs boson momentum is then identified
with the total momentum of two $b$-jets, it is not guaranteed that the
momentum of the $b\bar b$ dijet system equals the momentum of the
Higgs boson.  In principle, to reconstruct $b$-jets
properly, we need to carefully track flavor in the production process
since initial-state partons include $b$ quarks  that  propagate into the
final state, leading to $b$ jets that originate in  the production process. 
Also,
starting from NNLO final state splittings $g^*\to b\bar b$ 
generate additional  $b\bar b$ pairs in the final state. However, it can be
checked that both of these effects are rather small for WBF. Indeed, at
NLO  $b$-quarks in the
final state contribute about one percent of the cross section. 
Since we do not expect this result to change significantly at NNLO,
 we decided to  treat all quarks
 coming from the production stage as flavorless and not account for them when determining $b$-tags of
 jets. 

As a second approximation, we note that in this paper we consider
Higgs boson decays to massless $b\bar b$ pairs only at leading order
in QCD. It will be desirable to extend this calculation and include 
higher order QCD  corrections to Higgs decay as well.
We leave
this for future work but we note that the computation reported  in
this paper, that describes an interplay of NNLO QCD corrections to the production stage 
with leading order decays, is an important step towards this goal.

To account for Higgs boson decays, we make use of the fact that Higgs
bosons are narrow scalar particles, so that  their  production and decay
subprocesses are not correlated and can be considered separately.  For
this reason, we generate kinematics of final state particles in the $H
\to b \bar b$ process in the Higgs boson rest frame independently of
the production stage.  Then, we use high-quality importance-sampling
grids that describe stable Higgs production in weak boson fusion to generate properly distributed events with a given Higgs
four-momentum. For each production point, we consider ${\cal O}(10)$  randomly generated decay events. We then adjust the weight of each event to
account for the $H\to b\bar b$ branching ratio, which we take to be
${\rm Br}(H\to b\bar
b)=0.5824$~\cite{LHCHiggsCrossSectionWorkingGroup:2016ypw}.  
Finally, we
boost the momenta of $b$-quarks  back to the laboratory frame,
use them to reconstruct jets and check if final state objects pass
the selection criteria that we now describe. 

Similar to the case of a  stable Higgs boson, we reconstruct jets using the anti-$k_t$
algorithm and $b$-tag them  according to whether they contain a single $b$-quark
or not. As we have already said, we work
in an approximation where we have a single $b\bar b$ pair
coming from the Higgs so this procedure is infra-red safe. 
In addition to the kinematic cuts described in the previous section, we
loosely follow Ref.~\cite{ATLAS:2020bhl} and 
require that an event contains at least two $b$-jets with $p_{\perp j_b} \ge
65~{\rm GeV}$ and with their rapidities  confined to the interval
$|y_{j_b (j_{\bar b})}| \le 2.5$. The WBF cuts described in the
preceding section are applied exclusively to non $b$-jets.

We turn to the presentation of our results. We first report
values of fiducial cross sections.  Using
the setup described above, we obtain
\begin{equation}
\sigma_{\rm LO}^{b \bar b} = 75.9^{-5.6}_{+6.5}~{\rm fb},\;\;\;\;\;\;
\sigma^{b \bar b}_{\rm NLO} = 70.9^{+0.2}_{-1.2}~{\rm fb},\;\;\; \;\;
\sigma_{\rm NNLO}^{b \bar b} = 69.4^{+0.5}_{-0.2}~{\rm fb}.
\label{eq23}
\end{equation}
The central values correspond to the scale $\mu_0$ shown in
Eq.~(\ref{eq21}); the subscript and superscript show the cross section
at $\mu = \mu_0/2$ and at $\mu = 2 \mu_0$ respectively. 

It is interesting to compare the size of fiducial NLO and NNLO QCD
corrections for stable and decaying Higgs boson.  Using
Eqs.~(\ref{eq22},\ref{eq23}) we find (for $\mu = \mu_0$)
\begin{equation}
\begin{aligned}
\frac{\sigma^H_{\rm NLO}}{\sigma^H_{\rm LO}} &= 0.917(1) \, , \;\;\; & \frac{
  \sigma^{b \bar b }_{\rm NLO} }{ \sigma^{ b \bar b}_{\rm LO}} &=
0.934(1) \, , \\ 
\frac{\sigma^H_{\rm NNLO}}{\sigma^H_{\rm LO}} &= 0.885(1) \, ,\;\;\; &
\frac{ \sigma^{b \bar b }_{\rm NNLO} }{ \sigma^{ b \bar b}_{\rm
    LO}} &= 0.914(2) \, ,
\label{eq30}
\end{aligned}
\end{equation}
where the Monte Carlo integration error is shown in parentheses.
Eq.~(\ref{eq30}) shows that for stable Higgs boson the NNLO QCD cross
section is smaller than the leading order cross section by about $-11.5$
percent whereas for $b \bar b$ final state the NNLO QCD cross section
is smaller than the leading order cross section by about $-8.5$ percent. This
$3$ percent difference is clearly not a large one; however, it is
of the same order as the NNLO
QCD corrections themselves.  It is also interesting that the
difference between the two scenarios is slightly more pronounced at NNLO than
at NLO, where the corrections decrease the LO fiducial cross section
by about $-8$ percent for a stable Higgs and by about $-6.5$ percent when we allow for  Higgs  decays.

\begin{figure}[t]
  \centering
  \includegraphics[width=0.328\textwidth]{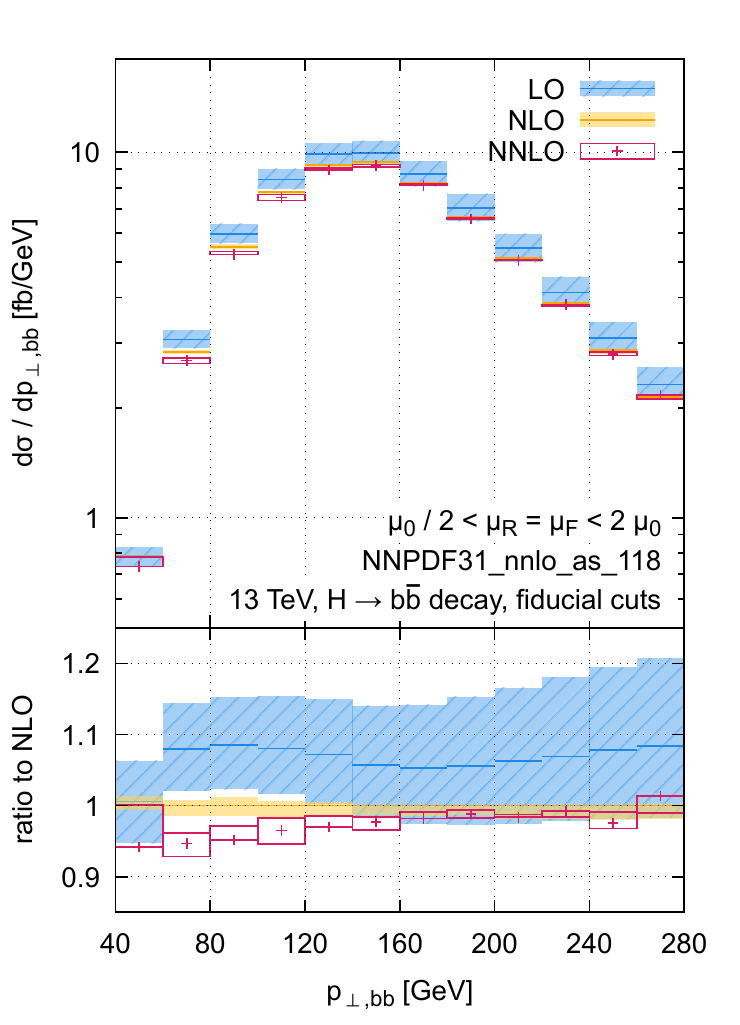}
  \includegraphics[width=0.328\textwidth]{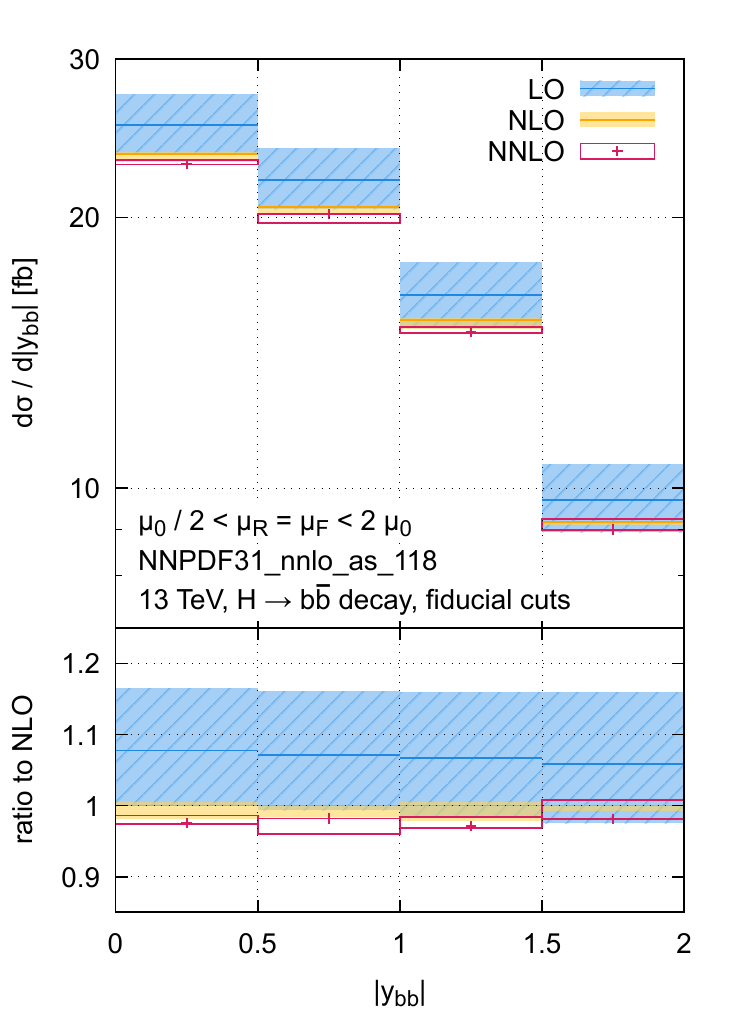}
  \caption{Transverse momentum (left) and rapidity (right) distribution
    of the reconstructed Higgs boson in the $H\to b\bar b$
    decay channel. See text for details.}
  \label{fig:hbb_higgs}
\end{figure}

As we now explain,  this  difference is largely caused by the cuts  on
the transverse momenta of the $b$-tagged jets, which make the Higgs
$p_\perp$ spectrum harder. This is illustrated in Fig.~\ref{fig:hbb_higgs}, where we show the transverse
momentum and rapidity distributions of the  $b \bar b$-dijet system that should be identified with the Higgs boson.   By comparing with the analogous plots in Fig.~\ref{fig::ptrap},
we see that the differential $K$-factors are rather similar, but now the
reconstructed Higgs transverse momentum peaks around 150~GeV instead of 
100~GeV which is the case if no cuts on the $b$-jets are  imposed, see
Fig.~\ref{fig::ptrap}.
This observation explains the difference in $K$-factors. 
Indeed, if we impose the requirement $p_{\perp,H} \ge 150$ GeV on our
results for the stable Higgs,  we find that the ratio of NNLO to
LO fiducial cross sections becomes about $0.91$, quite similar to what 
we find by  considering  $H \to b\bar b$ decays and imposing cuts on $b$-jets.
Hence, our results  show that a decent estimate
of the fiducial $K$-factor
in this case  can be obtained by considering    stable Higgs boson 
and computing  the $K$-factor  with the cut $p_{\perp,H}\ge 150~{\rm GeV}$.\footnote{We note
that it is customary to impose a $p_{\perp,b\bar b} \ge 150~\rm{GeV}$
cut in this channel~\cite{ATLAS:2020bhl}.}

Compared to the  results that do not include decays of the Higgs boson
presented in the previous section,
it is  interesting to note that in the $H\to b\bar b$ fiducial
region the NNLO QCD cross section overlaps with the scale-uncertainty band
of the NLO QCD  cross section. Furthermore, the relative scale variation of the NNLO result is
smaller by about a factor  two compared to the stable Higgs result.
These features are also explained by the fact that, when $H \to b \bar b$ decays
are considered, the Higgs  boson 
typically has larger  transverse momentum and, 
as follows from  Fig.~\ref{fig::ptrap}, at higher $p_\perp$ the NNLO result tends to get closer
to the NLO one  and  lies within the NLO scale variation band.

\begin{figure}[t]
  \centering
  \includegraphics[width=0.328 \textwidth]{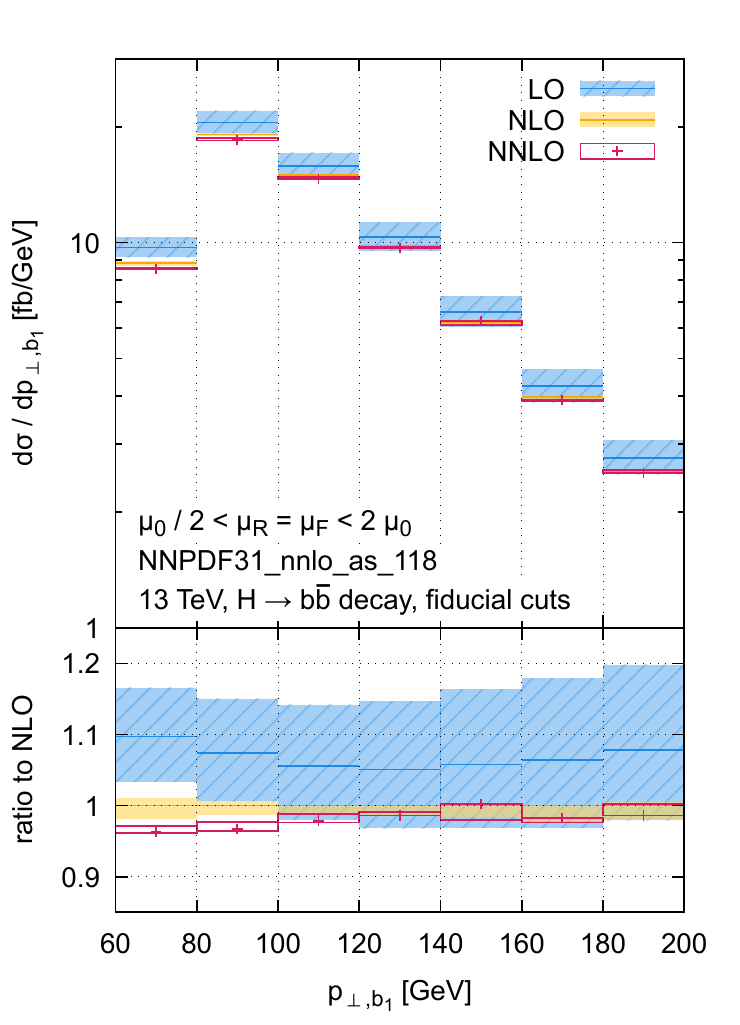}
  \includegraphics[width=0.328 \textwidth]{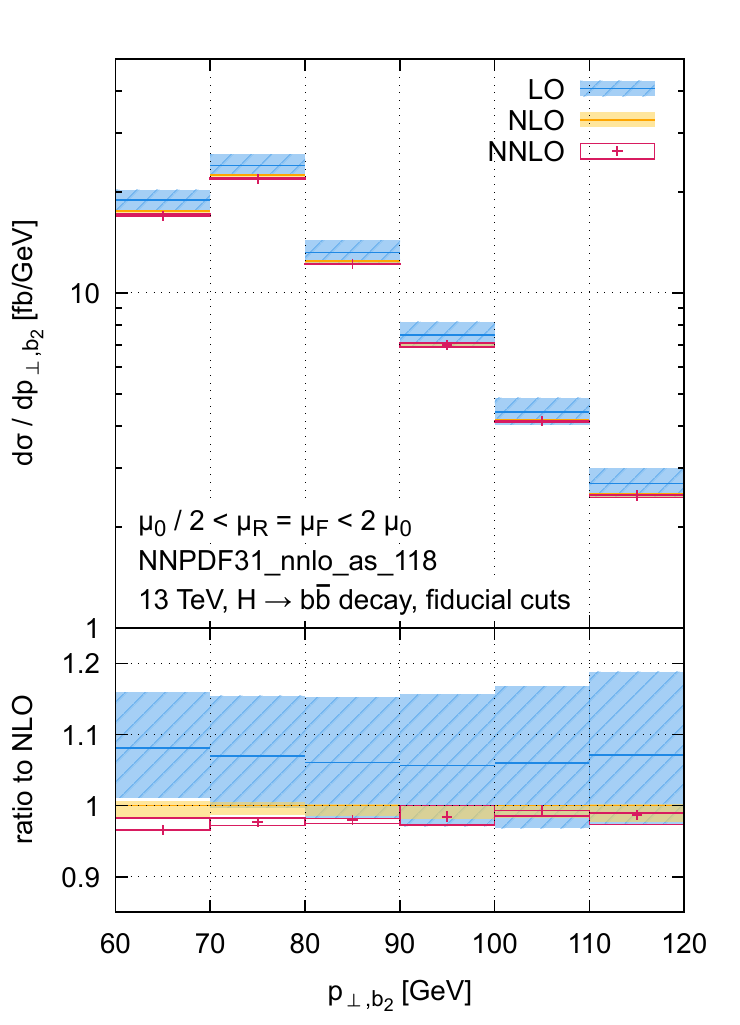} \\
  \includegraphics[width=0.328 \textwidth]{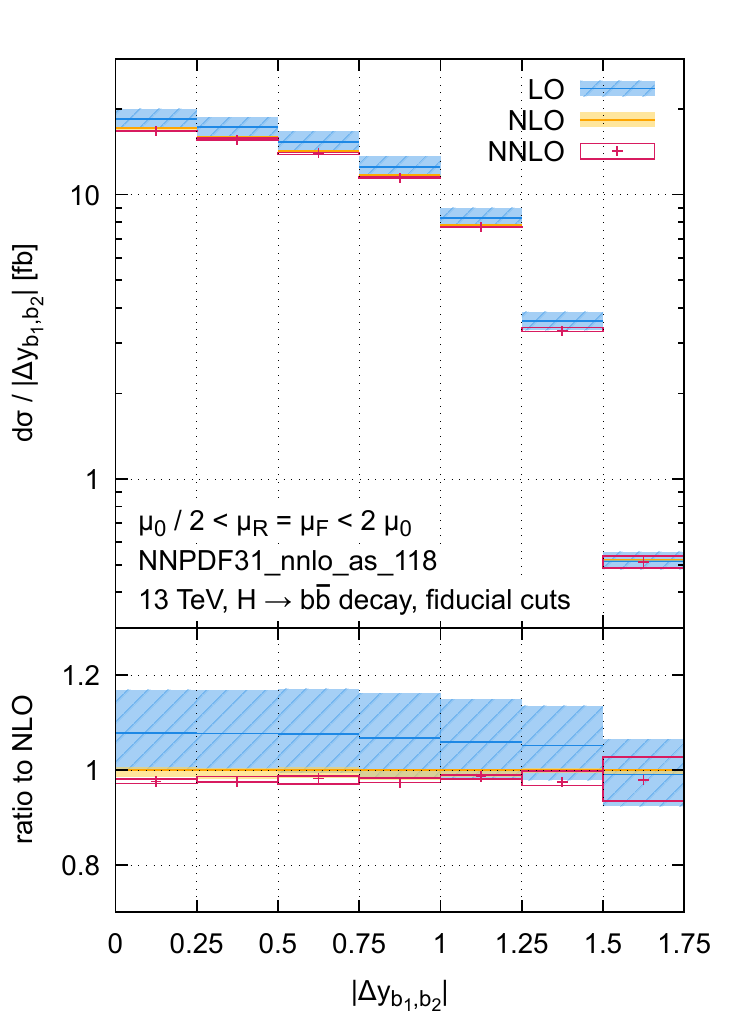}
  \includegraphics[width=0.328 \textwidth]{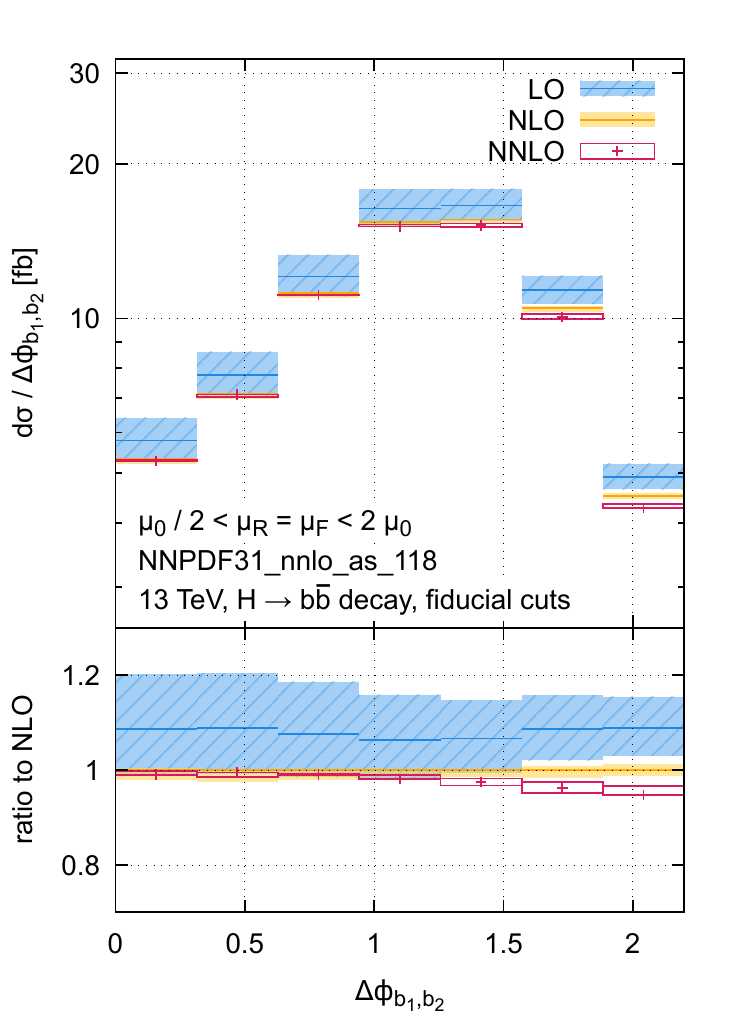}
\caption{Various kinematic distributions that involve $b$-jets from
  Higgs boson decays. Top row: leading (left) and subleading (right)
  $b$-jet transverse momentum distribution. Bottom row: rapidity (left)
  and azimuthal (right) separation between the two leading $b$-jets.
  See text for details.
}
\label{fig::bjets}
\end{figure}  

We continue with the discussion of kinematic distributions for
observables that involve the reconstructed $b$-jets.  They are shown
in Fig.~\ref{fig::bjets}, where we plot the transverse momentum
distribution of the leading-$p_\perp$ ($b_1$) and subleading-$p_\perp$ ($b_2$)
reconstructed  $b$-jets, as well as their rapidity
and azimuthal separation. We find that in the bulk of the distribution
the $K$-factors are rather flat, with the possible exception of the
leading-$p_\perp$ $b$-jet at NLO. Similar to the stable Higgs case,
the NNLO/NLO $K$-factor is flatter than the NLO/LO one.

\begin{figure}[t]
\centering
\includegraphics[width=0.558 \textwidth]{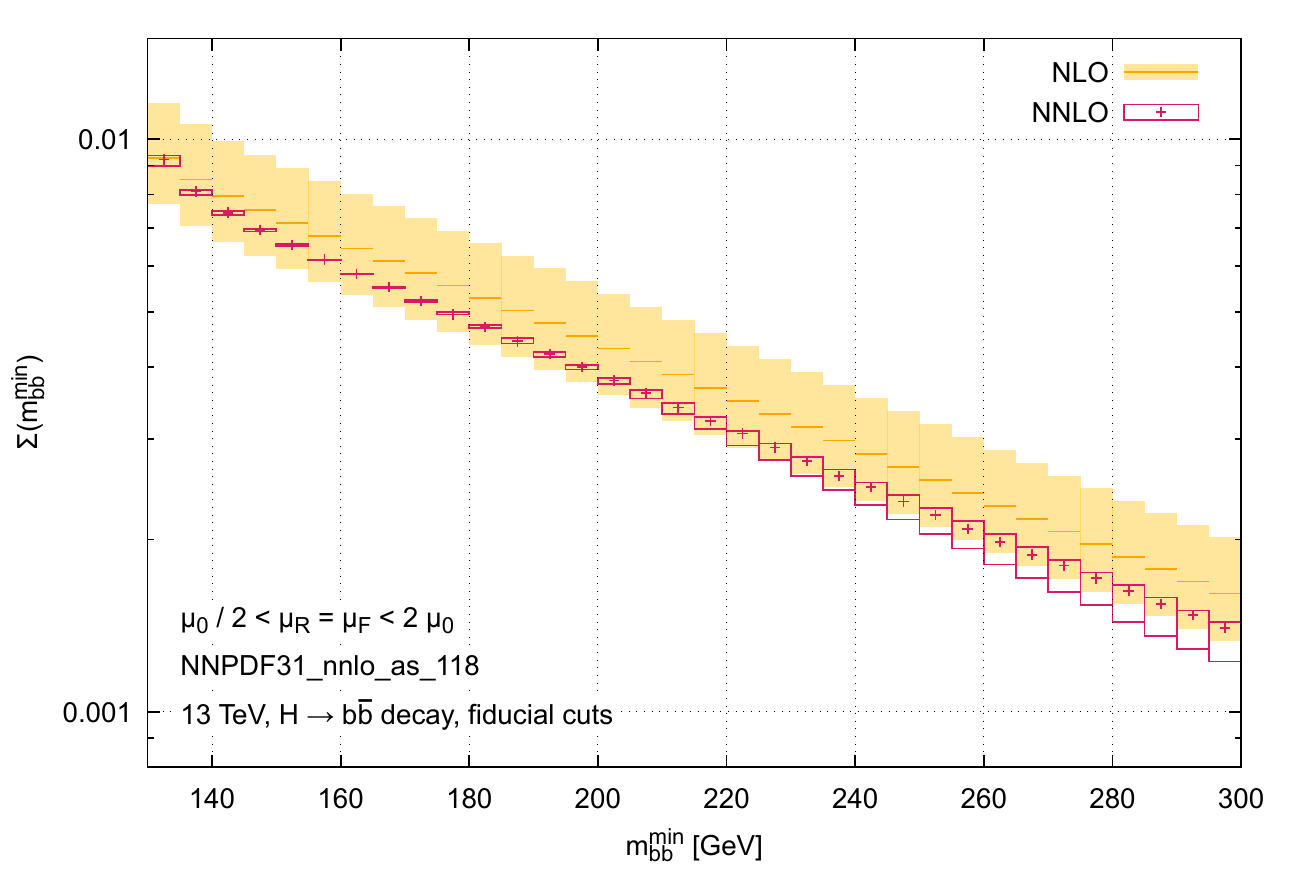}
\caption{Fraction of events $\Sigma$ for which the reconstructed Higgs mass is larger than the true Higgs mass. See text for details.}
\label{fig::mbb}
\end{figure}

As we mentioned earlier, $b$-quarks from Higgs decays may get recombined
with some of the partons in the production process.
If this happens, the reconstructed Higgs boson will have a non-trivial
invariant mass distribution, even if the Higgs boson is on-shell.
Since the Higgs
boson is produced centrally in WBF, while QCD radiation is mostly
collinear to leading jets, we expect this to happen very rarely.
This expectation is confirmed by Fig.~\ref{fig::mbb}, where we show
the fraction of events $\Sigma$ where the reconstructed Higgs mass $m_{b\bar b}$
is larger than a given value $m_{b\bar b}^{\rm min}$, i.e.
\begin{equation}
  \Sigma(m_{b\bar b}^{\rm min}) = \frac{1}{\sigma}
  \int\limits_{m_{b\bar b}^{\rm min}}^{\infty} {\rm d}m_{b\bar b}
  \frac{{\rm d}\sigma}{{\rm d}m_{b\bar b}}.
\end{equation}
In Fig.~\ref{fig::mbb}, we normalize the (N)NLO invariant mass distribution
to the corresponding fiducial cross section, integrated over the
invariant mass. 
We see that indeed only about 1\% of the events have a reconstructed
mass $m_{b\bar b}$ that exceeds the Higgs mass $M_H = 125$ GeV.

\subsection{Results for $H \to WW^* \to \ell^- \bar \nu \ell^+ \nu$}
\label{sect3c}

We now turn to Higgs production in weak boson fusion followed by the
Higgs decay to two $W$ bosons which decay leptonically, $H \to WW^*
\to e \bar \nu_e \bar \mu  \nu_\mu$.
Although this is a higher
multiplicity final state than considered in the previous section, from
a conceptual point of view it is simpler in the sense that the Higgs
decay products are not clustered with any of the QCD partons.
As done for $H \to b\bar b$ in the previous
section, we account for the $H \to WW^* \to  e \bar \nu_e \bar \mu  \nu_\mu $
decay by generating kinematics of the decay products in the Higgs rest frame 
and 
boosting  the decay products into the lab frame.  Compared to
the $H\to b\bar b$ case considered earlier, the decay phase space is
now much larger. To properly sample it, we then consider ${\cal O}(100)$
randomly
selected decay events per production point, instead of ${\cal O}(10)$  events
that we used for $H\to b\bar b$. In general, we note that, compared to $H \to b \bar b$,
it is much harder to efficiently sample  the phase space 
 in the four-lepton case.

We now define the fiducial region  for this decay channel.    In addition to
the cuts described in Section~\ref{sect3a}, we impose cuts on the 
particles  arising from the Higgs decay following 
Ref.~\cite{CMS:2018zzl}. We require that the leading charged lepton has
transverse momentum $p_{\perp,l_1} \ge 25 $ GeV while the subleading charged 
lepton should have transverse momentum $p_{\perp,l_2} \ge 13 $ GeV. The
invariant mass of the charged-leptons system should satisfy $m_{l_1
  l_2} \ge 12$ GeV, its transverse momentum $p_{\perp}^{l_1l_2}$ should exceed $30$~GeV, the missing transverse momentum
should be larger than $20~{\rm GeV}$ and 
the rapidities  of the charged leptons should be
between the rapidities of the two hardest jets. We note that this cut
correlates the production and decay stages and, given the forward nature of leading jets, it selects  Higgs bosons  produced in the
central rapidity region. 
Finally, we require the
transverse mass, defined as
\begin{equation}
m_T = \sqrt{ 2 p_\perp^{l_1 l_2} p_\perp^{\rm miss}\left( 1 - \cos \Delta
  \phi_{l_1 l_2, \vec{p}_{\perp}^{\rm miss}} \right) }, 
\label{eq:mT}
\end{equation}
to satisfy $60~\mathrm{GeV} \le m_T \le 125$ GeV. In Eq.~(\ref{eq:mT}),
$p_T^{l_1 l_2}$ and $p_T^{\rm miss}$ are the transverse momenta of the
charged-leptons system and the missing transverse momentum,
respectively, while $\Delta \phi_{l_1 l_2, \vec{p}_{T}^{\rm miss}}$ is
the azimuthal angle between the transverse momenta of the charged-leptons and two-neutrinos systems.

Computing fiducial cross sections, we obtain 
\begin{equation}
\sigma_{\rm LO}^{e \bar \nu_e \bar \mu  \nu_\mu} =
0.719^{-0.045}_{+0.051}~{\rm fb},\;\;\;\;\;\; \sigma^{e \bar \nu_e \bar \mu  \nu_\mu}_{\rm NLO} = 0.662^{+0.005}_{-0.012}~{\rm fb},\;\;\; \;\;
\sigma_{\rm NNLO}^{e \bar \nu_e \bar \mu  \nu_\mu} =
0.632^{+0.008}_{-0.008}~{\rm fb}.
\label{eq:res4l}
\end{equation}
The pattern of the corrections is very similar to what we observed in
Section~\ref{sect3a} for the stable Higgs case: the NLO corrections reduce the
cross section by $-8$ percent and the NNLO corrections reduce it by additional
$-4.5$ percent. The relative scale variation uncertainty is
also similar to the stable Higgs case. These features can be understood since, in contrast to Section~\ref{sect3b}, we impose relatively mild
cuts on the Higgs decay products which do not force the kinematics of the 
Higgs boson to differ significantly from the stable case.   Indeed, we have checked that, in this fiducial
region, the differential distributions of 
 the Higgs and the jets   are very similar to the corresponding results for the stable Higgs boson.
  
\begin{figure}[t]
  \centering
  \includegraphics[width=0.328\textwidth]{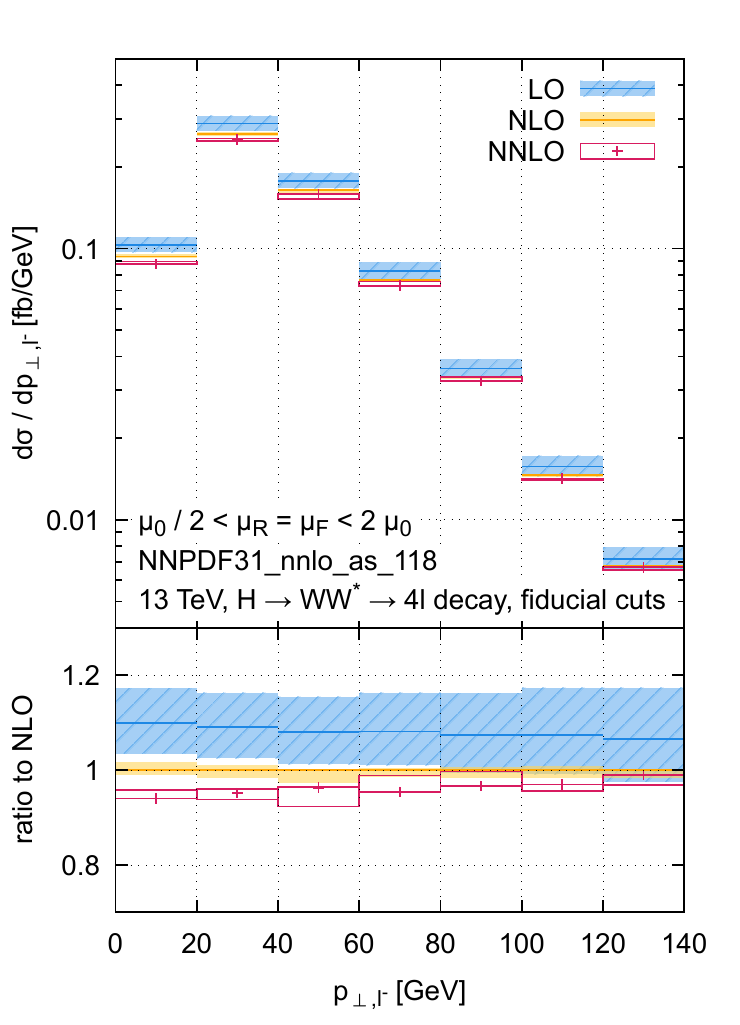}
  \includegraphics[width=0.328\textwidth]{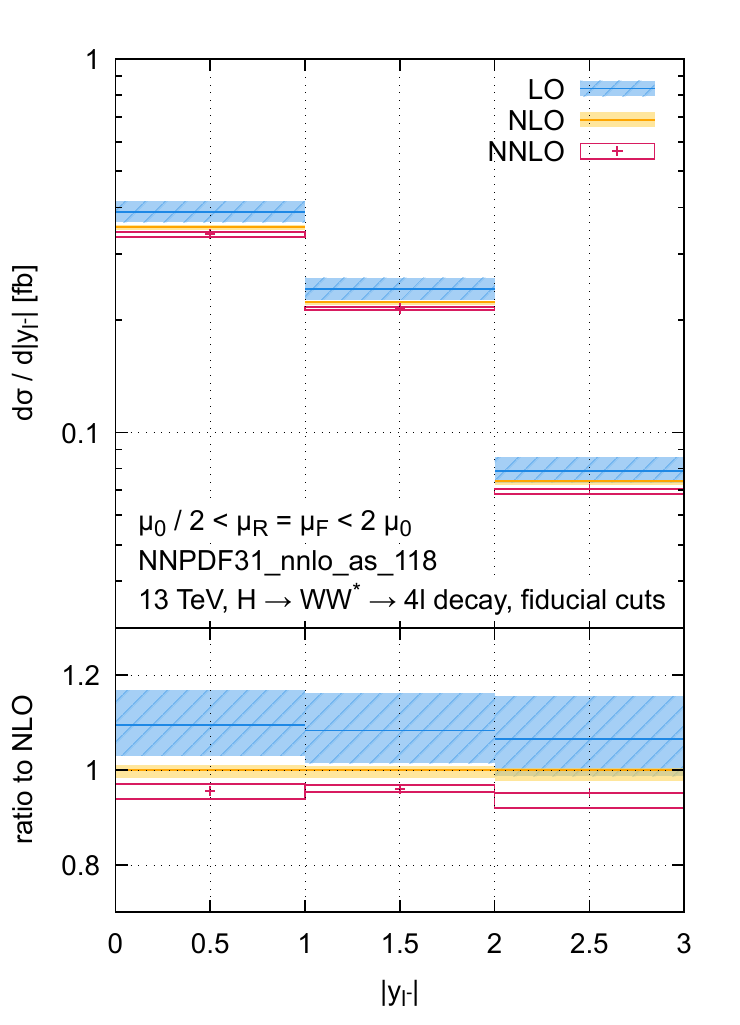}
  \includegraphics[width=0.328\textwidth]{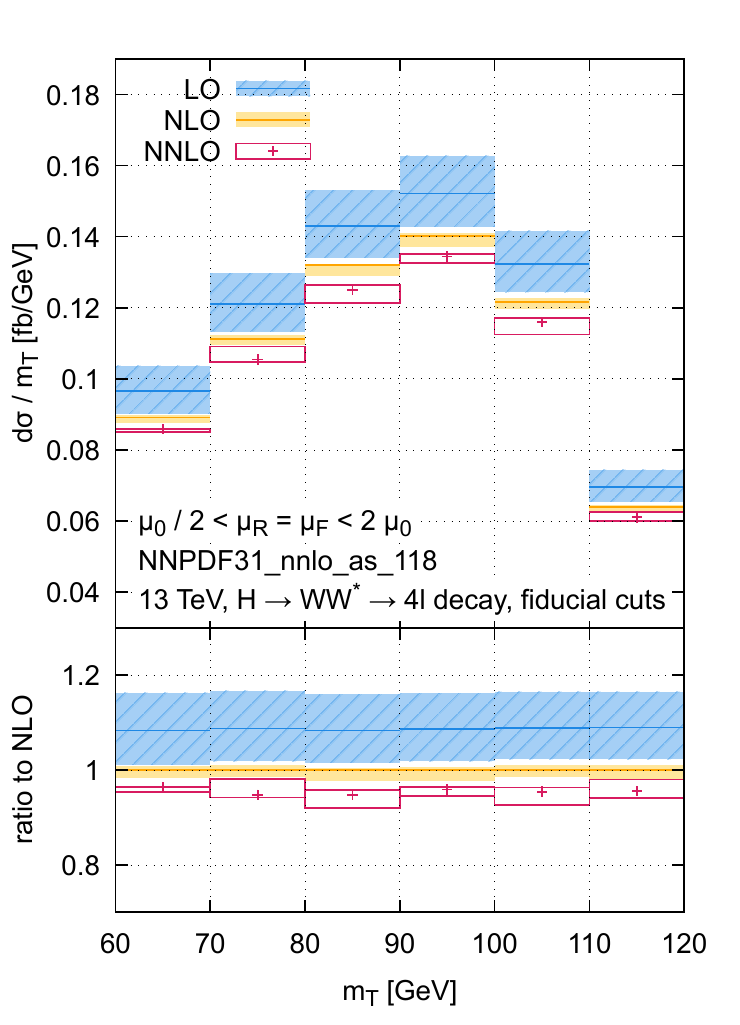}
  \caption{Various kinematic distributions that involve charged
    leptons from Higgs boson decays. From left to right: transverse
    momentum and rapidity distribution of the negatively charged
    lepton, and transverse mass of the charged-leptons system. See
    text for details. }
\label{fig::leptons}
\end{figure}

In Fig.~\ref{fig::leptons} we present results for selected kinematic
distributions of the two charged leptons. We show the
transverse momentum and rapidity of the negatively-charged lepton and the transverse
mass defined in Eq.~(\ref{eq:mT}).
We see that already at NLO the $K$-factors are  rather flat and this remains  true at NNLO. This is not
surprising since in this case the only impact of radiative corrections  comes from the interplay
between jet and lepton  cuts. As we mentioned earlier, the leptonic
cuts that we employ are rather mild and do not severely affect the kinematics of the Higgs boson. In particular,
the only cut that correlates jets and leptons is the requirement that the charged lepton rapidity should
lie within the rapidities of the two hard jets. However, the tagging jets are mostly produced along the beam line, 
so this requirement is satisfied by most of the events.

In summary, our results show that in this channel the impact of Higgs decays on radiative
corrections is milder. This happens because the kinematic features  of the Higgs boson and of the  jets 
remain  unaffected by the fiducial cuts on leptons  and, as a consequence,  corrections to leptonic observables are
rather flat and can be described to a very good approximation with a global $K$-factor corresponding
to fiducial cross section for stable Higgs. 

\newpage
\section{Conclusions}
\label{sect4}
In this paper, we presented a computation of NNLO QCD corrections to
Higgs boson production in weak boson fusion using the nested
soft-collinear subtraction scheme.  We have used analytic formulas for
the required NNLO integrated subtraction terms derived in
Ref.~\cite{Asteriadis:2019dte}.  We have shown that, although we
employ a somewhat different phase-space parametrization for the current
computation, the results derived in
Ref.~\cite{Asteriadis:2019dte} remain applicable.

We have confirmed earlier results on NNLO QCD corrections
to fully-differential Higgs boson production in weak boson fusion
obtained in Refs.~\cite{Cacciari:2015jma,Cruz-Martinez:2018rod}.  We
have also extended these results by incorporating  decays of the Higgs
boson into the calculation. We considered two Higgs  decay modes that are important
for WBF
studies, namely $H \to b \bar b $ and $H \to WW^* \to 4$~leptons.  We observed 
that in the $H\to b\bar b$ case the perturbative behavior of the
fiducial cross section differs from the stable Higgs case.  In
particular, we found that, while the NLO/LO ratio is very similar for
stable Higgs and for $H\to b\bar b$, the difference in the
NNLO/LO ratio is comparable to the  NNLO QCD corrections themselves.

We have
argued that the main reason for this difference  is that the cuts on
the $b$-jets push the transverse momentum of the Higgs boson towards larger
values, where NNLO corrections are small. If this effect is taken into
account, the impact of radiative corrections in the stable-Higgs
approximation and in the $H\to b\bar b$ decay channel become very
similar. In both cases, shapes of NLO distributions are not significantly affected  by NNLO corrections so that  a  rescaling of NLO distributions
would provide a good approximation to the
full result. We stress that this would \emph{not} be the case if one
were to use a NNLO/NLO $K$-factor computed in the standard WBF
fiducial region, without an additional $p_{\perp,H}\gtrsim 150~{\rm GeV}$
cut. 

In the $H\to WW^*\to 4l$ channel, typical selection cuts are milder
and do not significantly affect the Higgs and jets kinematic distributions.
Because of this, corrections to Higgs and leptonic observables are rather flat
and can be well captured by an overall $K$-factor. 

Our results could be extended in several directions. Most prominently,
our $H\to b\bar b$ analysis is only approximate. Indeed, we did not
consider radiative corrections to $H\to b\bar b$ decay. Given the non-trivial
interplay between jets coming from the production stage and jets originated
from the $b$-quarks from   Higgs decays, it would be interesting to
perform a \emph{complete} NNLO analysis that accounts for corrections
to both the production and decay stages of the WBF process.

Furthermore, since  radiative corrections to WBF in the fiducial region
are impacted by a non-trivial jet dynamics, it is difficult 
to predict how potential Beyond the Standard Model effects would impact the radiative
corrections. To study this point, one could, for example, repeat the NNLO QCD calculation within the Standard Model effective field theory
framework, to investigate to which extent radiative corrections can mimic
potential anomalous couplings effects if the latter are only predicted
at low orders in QCD perturbation theory. We leave these interesting
avenues of investigation for the future. 

{\bf Acknowledgments:}
We would like to thank Alexander Karlberg for helping us with
comparisons with the results of Ref.~\cite{Cacciari:2015jma} and for many
interesting discussions. We also thank Tobias Neumann for helpful advice
about \texttt{MCFM}.  
This research is partially supported by the Deutsche
Forschungsgemeinschaft (DFG, German Research Foundation) under grant
396021762 - TRR 257. The research of F.C. was partially supported by
the ERC Starting Grant 804394 {\sc hipQCD} and by the UK Science and
Technology Facilities Council (STFC) under grant ST/T000864/1. The research
of K.A. is supported by the United States Department of Energy under Grant Contract
DE-SC0012704.

\appendix

\bibliography{wbf}{}

\end{document}